# Retail prices, environmental footprints, and nutritional profiles of commonly sold retail food items in 181 countries

Elena M. Martinez, PhD [1], Nicole Tichenor Blackstone, PhD [1 2], Parke E. Wilde, PhD [1], Anna W. Herforth, PhD [3], William A. Masters, PhD, FASN [1 4]

[1] Friedman School of Nutrition Science and Policy, Tufts University, USA
[2] Department of Biomedical Engineering, Tufts University, USA
[3] Food Prices for Nutrition Project, Boston, USA
[4] Department of Economics, Tufts University, USA

Correspondence: Elena M. Martinez, elena.martinez@tufts.edu

**Acknowledgement:** This work was conducted as part of the first author's PhD dissertation at the Friedman School of Nutrition Science and Policy, Tufts University, supported by the Gates Foundation and the UK government through the Innovative Methods and Metrics for Agriculture and Nutrition Actions project (INV-002962) and the Food Prices for Nutrition project (INV-016158), with additional support from the World Bank and the Food and Agriculture Organization of the United Nations. Related work on food prices and diet costs is here: https://sites.tufts.edu/foodpricesfornutrition.

**Abstract**

**Background:** Transitions towards healthier, more environmentally sustainable diets would require large shifts in consumption patterns. Cost and affordability can be barriers to consuming healthy, sustainable diets.

**Objective:** This study provides the first worldwide test of how retail food prices relate to empirically estimated environmental footprints and nutritional profile scores between and within food groups.

**Methods:** We use 48,316 prices for 860 retail food items commonly sold in 181 countries during 2011 and 2017, matched to estimated carbon and water footprints and nutritional profiles, to test whether healthier and more sustainable foods are more expensive between and within food groups.

**Results:** Prices, environmental footprints, and nutritional profiles differ between food groups. Within almost all groups, more expensive items have significantly larger carbon and water footprints. Associations are strongest for animal source foods, where each 10% increment in price is associated with 21 grams higher carbon footprint and 5 liters higher water footprint per 100kcal of food. There is no such gradient for price and nutritional profile, as more expensive items are sometimes healthier and sometimes less healthy depending on the food group, price range, and nutritional attribute of interest.

**Conclusions:** Our finding that higher-priced items have larger environmental footprints is contrary to expectations that a more sustainable diet would be more expensive. Instead, we find that within each food group, meeting dietary needs with lower environmental footprints is possible by choosing items with a lower unit price. These findings are consistent with prior observations that higher-priced items typically use more resources, including energy and water, but may or may not be healthful as measured by nutrient profile scores.

**Abbreviations:** ICP: International Comparison Program; $CO_2$eq: carbon dioxide equivalent; ASF: animal source food; CF: carbon footprint; WF: water footprint; FCS: Food Compass Score

## Introduction

Food systems are significant contributors to environmental crises including climate change, water scarcity, biodiversity loss, and pollution. Food systems account for as much as one third of anthropogenic greenhouse gas emissions (1) and 70 percent of freshwater withdrawals (2), as well as approximately 32 percent of terrestrial acidification and 78 percent of aquatic eutrophication (3). Current diets are high in unhealthy and environmentally harmful foods, and many populations suffer from both food insecurity and diet-related diseases (4-6).

Transitions towards healthier, more environmentally sustainable diets would require large shifts in consumption between and within food groups (4,7), but prices and affordability can constrain food choice (5). Even the least expensive healthy diets are not yet affordable for nearly three billion people globally (6). Many national food-based dietary guidelines, as well as the EAT-Lancet reference diet, are similarly unaffordable (9-11). Consumers and policy analysts frequently state that more sustainable or healthier foods are systematically more expensive than conventional choices (12,13), in part because market prices do not account for the external costs of unsustainable or unhealthy foods (14-17). When producers adhere to standards designed to improve sustainability and health, they often incur higher costs of production (18-21), and these products are sold at higher retail prices than conventional options (22-24). Individuals with higher diet quality scores also often have higher food expenditure per day in settings as diverse as Belgium, Mexico, and the United States (25-27). Those observations all imply that transitioning to more sustainable, healthier diets would require higher consumer expenditure, but other data suggests that sustainability and health could be improved at lower cost to consumers, such as by replacing beef and other expensive animal-source foods with nutrient-rich plant-based foods (28-31).

This study provides the first global test across all major food groups of how item prices from retail food outlets around the world relate to the food item's environmental footprints and nutritional value. We build on past work that focuses on price per calorie across food groups globally (32). We combine retail food prices from 181 countries with estimates of the carbon footprint, water footprint, and nutritional profile of these food items to assess whether more environmentally sustainable and healthier foods are more expensive than other options. Identifying which foods are actually less expensive, within and between food groups, could inform interventions that achieve environmental and health goals at lower cost to consumers and society.

## Methods

We use the International Comparison Program (ICP) (33) global and regional datasets for 2011 and 2017, which provide national average prices and availability of 643 food items in 177 countries in 2011 and 545 food items in 175 countries in 2017, for a total of 48,316 price observations reported in Table 1. We convert prices from local currency to 2017 USD using purchasing power parity (PPP) exchange rates for individual consumption expenditure by households, provided by the ICP, excluding 5 countries/territories (Anguilla, Bonaire, Cuba, Montserrat, and Taiwan) for which PPP exchange rates were unavailable. We convert prices from 2011 to 2017 USD using inflation as measured by the consumer price index (34). We convert prices per unit as purchased (e.g., one dozen eggs) to prices per kilocalorie using information from ICP food item descriptions and food composition tables (35-37). (See Appendix 1a for equations.)

To compare prices between and within food groups, we classify items using the Healthy Diet Basket (HDB) used by United Nations agencies and the World Bank to monitor the cost and affordability of a healthy diet globally (9, 38). Based on commonalities among national food-based dietary guidelines, the HDB specifies six food groups needed for health (starchy staples; animal source foods; legumes, nuts and seeds; vegetables; fruits; oils and fats). For this paper, we also consider a discretionary category for sugars, sweets, and candies. We calculate price per 100 kcal, and price per day based on recommended daily intake of each food group in the HDB. (See Appendix 1a for equations.) To focus on comparisons between the main foods and food groups needed for healthy diets, we exclude beverages (except for fruit juices), culinary ingredients, spices, herbs, condiments, mixed dishes with unclear composition, and infant foods. When used to calculate the official cost and affordability of a healthy diet metrics, the HDB food groups exclude items generally not recommended as part of a healthy diet, such as processed meat and vegetable ghee. We include these items in this analysis to examine price differences by nutritional value within food groups.

For environmental footprints, we match food items to their associated carbon and water footprints using a recent, peer-reviewed database of estimates from around the world (39,40). These data provide global average estimates of carbon footprint, which represents the sum of cradle-to-retail gate greenhouse gas emissions and removals, in kilograms of carbon dioxide equivalent ($CO_2$eq) per kilogram of food for 324 food items, and water footprint in liters per kilogram of food for 320 food items based on the Global Water Footprint Standard, which combines blue, green, and grey water footprints (41-43). Global estimates are derived from a total of 3,349 individual carbon footprint estimates and 938 water footprint estimates following a rigorous, transparent process described by Petersson and colleagues (2021). We then use food composition data to convert both environmental footprints into quantities per kilocalorie and per day of recommended intake. (See Appendix 1c for detailed methodology for matching carbon and water footprints to food items.)

For nutritional values, we use three different metrics: Food Compass Score (44-46), which rates each food on a scale from 0 to 100 based on nine domains associated with health outcomes; Nutri-Score (47) from 0 to 5 based on nutrients to promote and to limit; and Health Star Rating (48) from 0.5 to 5 based on nutrients associated with chronic disease and health outcomes. Scores were assigned to the food items for which we have price data by matching item descriptions in our ICP database to items listed in the USDA Food and Nutrient Database for Dietary Studies (FNDDS). (See Appendix 1d for further details on Food Compass, Nutri-Score, and Health Star Ratings.)

In the dataset used to generate our results, each data point is the global average carbon footprint, water footprint, and nutrient profile score of a food item, matched to its national average retail price in every country where it was reported to be widely sold in 2011 or 2017.

(See Appendix 1e for data inclusion flow chart.) **Table 1** provides descriptive statistics of price, environmental footprints, and nutrient profile by food group. **Figures 1 and 2** show the relationship between price and carbon footprint, water footprint, and nutrient profile both within and between food groups using binned scatterplots, which summarize the large number of observations in centiles of their distribution. In these figures, thousands of observations in each food group are summarized using 100 data points, which represent the mean value of the y-axis variable for each of 100 equal-sized bins of item price observations along the x-axis. This type of data visualization reduces visual clutter and reveals variation in the attribute of interest at each level of price. We show the natural log of food prices for visual clarity because the distribution of food item prices is heavily skewed right. We selected the natural log transformation by using the Stata gladder command to identify which transformation converted item prices into a normal distribution.

We also test for the association between each attribute and food item prices using the following ordinary least squares regression model:

$$Y_{ij} = \beta_0 + \beta_1 * \ln(price)_{ij} + \alpha_i + \varepsilon_{ij}$$

In this regression, *Yij* is the attribute of interest (carbon footprint, water footprint, or nutrient profile score) in each country (*i*) for each food item (*j*), in natural units (kilograms of $CO_2$eq emissions, liters of water, or Food Compass Scores from 0 to 100, Nutri-Score from 0 to 5, Health Star Ratings from 0.5 to 5, or each of the 9 component domain scores used to calculate Food Compass Score from -10 to 10), reported either per 100 kcal or per daily recommended intake for items in that food group, and price is the retail price of each food item (*j*) in either PPP dollars per kcal or per daily recommended intake in each country (*i*). All regression results include country fixed effects and are stratified by food group, executed in Stata SE 18. Regression results are visualized in **Figures 3 and 4**.

**Results**

We find that more expensive food items have greater environmental footprints in almost all food groups and price ranges. While average environmental footprints vary between food groups, within each food group items with higher prices per 100 kcal or per day have larger carbon and water footprints than less expensive items in that food group. The relationship between price and nutritional profile score is more variable, as more expensive items are sometimes healthier and sometimes less healthy than less expensive items within the same food group.

*Carbon footprints and retail food prices*

Carbon footprints (CFs) per unit of dietary energy vary widely across food groups, with animal source foods (ASFs) having the highest CF per kcal at all price points (**Table 1**). ASFs also have the steepest gradient of CF with respect to price, and the next steepest gradient is for vegetables (**Figure 1**). Less expensive vegetables such as carrots and onions have similar CFs as other plant foods, while the most expensive vegetables such as asparagus and mushrooms have CFs as high or higher than the least expensive ASFs such as fresh small fish and eggs. The energy required from each food group to meet dietary recommendations differs greatly, however, and the larger recommended intakes of starchy staples and ASFs makes their CFs much larger than those of vegetables at all price points (**Figure 2**). Within each food group except starchy staples, more expensive food items have significantly larger CF per kcal or per day, with the average gradient for ASFs more than twice the magnitude of the gradient for vegetables (**Figure 3**). A 10% lower price is associated with 21 grams less $CO_2$eq per 100 kcal of ASFs, 8.3 grams less $CO_2$eq per 100 kcal of vegetables, and smaller gradients for the other food groups (Figure 3).

*Water footprints and retail food prices*

Like CFs, water footprints (WFs) are generally larger at higher prices for most food groups, though WF varies more than CF at each price point (Figures 1 and 2). ASFs have the highest WF per kcal and per day compared to other food groups, but the next highest WFs per kcal are for fruits and legumes, nuts and seeds (Figures 1 and 2). Less expensive vegetables have among the lowest WFs per kcal of any food group (Figure 1). Since dietary recommendations call for lower dietary energy contribution from vegetables than other food groups, vegetables have by far the lowest WF per day (Figure 2). WFs per kcal and per day are higher for higher-priced items in all food groups except starchy staples, and the magnitude of this association is largest for ASFs; legumes, nuts, and seeds; vegetables; and fruits. A 10% higher price is associated with 4.8 liters higher WF per 100 kcal of ASFs, 4.5 liters higher WF for legumes, nuts, and seeds, 3.6 liters higher WF for vegetables, and 3.4 liters higher WF for fruits (Figure 3). For ASFs, the gradient of WF with respect to price plateaus and reverses at higher price points. While some relatively expensive ASFs have high WF (e.g., beef), some of the most expensive ASFs in each country have comparatively low WF (e.g., some cheeses, fresh fish) (Figure 1).

*Nutritional profile scores and retail food prices*

Nutritional profiles are less consistently associated with price than either CF or WF, as shown using Food Compass Score (FCS) in the main text and two other nutrient profiling systems in the supplementary materials (Supplementary Figures 4a-4i). Fruits and vegetables have the highest FCS at all price points, followed by legumes, nuts, and seeds. Within each of these food groups, there is very little correlation between FCS and price, while within other groups there are steep gradients of FCS by price. For example, the least expensive oils (e.g., sunflower oil) and the most expensive ASFs (e.g., crab, shrimp) both have FCS similar to the lowest scoring fruits (e.g., orange juice) (Table 1, Figure 1). The magnitude of association between FCS and price is relatively small for all food groups, but for mid-priced ASFs there is a sufficiently steep gradient for a 10% higher price to be associated with a 0.98-point higher FCS (Figure 3). FCS varies little among less expensive ASFs, but in the most expensive quartile of ASFs there is large variation (e.g., ham has low FCS, while fresh snapper has high FCS) (Figure 2). Among oils and fats, the lowest price vegetable oils have high FCS, while more expensive options such as butter, ghee, and margarine have much lower FCS. On average, a 10% higher price for oils and fats is associated with a 1.9-point lower FCS (Figure 1, Figure 3).

The relationship between nutritional value and price differs by domain of the FCS, with positive gradients between FCS and price for some nutritional attributes, but negative gradients for others (**Figure 4**). In some food groups, nutrient ratios (between unsaturated and saturated fats, fiber and carbohydrates, and potassium and sodium) are more favorable in more expensive items, for example because whole grain products (e.g., whole wheat breads) are often more expensive than corresponding items made with refined grains. In the fats and oils category, however, nutrient ratios are more favorable in less expensive items, because plant oils high in unsaturated fats tend to be less expensive than animal products with more saturated fat (Figure 4, top left). Additives and phytochemicals are not strongly associated with price for any food group. Among vegetables, less expensive items are lower in minerals, vitamins, and fiber on average, while fruits have a smaller correlation between price and those nutrients (Figure 4). Results using Nutri-Score and Health Star Ratings are generally consistent with these results based on FCS, with some notable between-group differences in nutritional profile because the criteria used by these three nutrient profiling systems score some ASFs and starchy staples differently (Appendix 4).

**Discussion**

We combine worldwide data on retail food prices with each item's carbon footprint, water footprint, and nutritional profile score. We find that more expensive retail items within each food group generally have larger environmental footprints but do not always have a healthier nutritional profile. This result implies that dietary shifts towards lower environmental footprints could be achieved while significantly lowering cost to consumers, while the cost of shifting towards healthier options depends on which nutritional attributes are improved.

Past studies of transitions towards more sustainable, healthier dietary patterns have focused on differences between food groups, primarily comparing animal source foods to fruits, vegetables, and other plant-based foods (26-29). Our results are consistent with prior findings that most animal source foods have both higher retail prices and higher environmental footprints than most plant-based foods, and most fruits and nuts have higher water footprints compared to other plant-based foods (4). Our contribution is a systematic observation of environmental footprints and nutritional profiles of food items at all price points, revealing implications for food choice, policy priorities, and investments within and between food groups.

Our central finding is that carbon and water footprints are positively associated with price across the range of items in most food groups, while the associations between nutritional profile scores and price vary between food groups and across the range of items available in each food group. Food prices reflect the balance of supply, demand, and trade for each food product at each location, with higher prices for items for which supply chains use more expensive resources, such as land, feed for livestock, labor, and equipment, each of which has varying implications for greenhouse gas emissions and water use (49). Our finding that carbon and water footprints are larger for higher-priced items in each food group is based on global average environmental footprints, reflecting the mix of carbon and water footprints observed in thousands of analyses for hundreds of products in countries around the world. We find variation around the trend, but the strong positive correlation between price and environmental footprint reflects the use of resources to meet consumer demand for food attributes such as taste, convenience, cultural preferences, and aspirations.

In contrast, the ingredients and processing techniques used to make higher-priced items in each food group sometimes raise and sometimes lower the item's nutritional profile score. Variation in the link between price and nutritional attributes can be illustrated by comparing trends for different food groups. Among oils and fats, butter and ghee are more expensive and have lower health scores than plant oils due to their high ratio of saturated to unsaturated fats. In contrast, among vegetables and fruits, more expensive items such as asparagus and berries tend to have a higher density of vitamins, minerals, and known phytochemicals than less expensive options. Among starchy staples, more expensive items such as unsweetened breakfast cereals and white bread tend to have fewer unhealthy ingredients but also more unhealthy forms of processing than less expensive options. This finding shows that consumers' stated beliefs that healthier foods are more expensive (12) is only sometimes consistent with the evidence. Despite the fact that many consumers can and do pay a premium for products that are advertised as healthier (22-24), lower-priced options within each food group are often equally or more nutritious.

Our results imply that choosing lower-priced options within each food group will, on average, lower environmental footprints, with varied impacts on diet quality. However, the environmental footprints and nutritional profiles of foods are not directly visible at retail. Thus, providing consumers with accurate and useful information about the composition and footprints of each item can help guide food choice, complementing other attributes such as price, appearance, and taste. Our results confirm the need for food labeling systems that convey information about each item's impact on both the environment and human health, such as the Eco-Score system introduced alongside Nutri-Score in some European countries (50). Labeling systems can disclose environmental footprints and nutritional profiles in many different ways, using summary statistics per 100 kcal or relative to quantities recommended per day, as done in this study. These disclosures could potentially be combined with mandatory disclosure of standardized unit prices (e.g., per 100 grams), which has been shown to help consumers choose among different package sizes for the same item (51), much as disclosure of health effects influences food choice independently of price (52). Combining the two could help consumers decide what they are willing to pay for each item, given its price per unit, environmental footprint, and nutritional quality.

*Strengths and limitations*

This is the first global analysis connecting retail food prices to the environmental footprints and healthfulness of items sold in retail markets worldwide. The study's strengths include using the most complete available set of nationally representative food prices (33), matched to a comprehensive database of global average carbon and water footprints from life cycle analyses around the world (39,40), and three different nutritional profiling systems created in the United States, Europe, and Australia to score the healthfulness of each food item (44-48). We match item descriptions for each food in each country to the available food composition data (35-37) and convert prices and environmental footprints to standardized units per 100 kcal and per day of the global reference diet used to measure food access by UN agencies and the World Bank (6,10). We then show the relationship between price and environmental footprints or nutritional profiles using direct observation in binned scatterplots, where each point represents one percent of the entire sample for each food group, and ordinary least squares regression, where each coefficient represents the correlation of interest with statistical controls for country and food group. This study design is intended to provide the clearest possible data visualization of how environmental footprints and nutritional profile scores vary with product price around the world.

Our research question concerns patterns observed on average across the world. One key limitation is that environmental footprints vary between geographic locations and types of food systems. We use global averages derived by Petersson et al. (2021) from based on a comprehensive review of available literature (39). These global averages are calculated from 4,287 footprint estimates (3,349 carbon footprints and 938 water footprints), conducted in 82 countries on 6 continents. The representativeness of this database is skewed by the available estimates of environmental footprints from past studies, over-representing estimates from wealthier countries and certain food products, for which more analyses are conducted. Carbon and water footprints were available for only 78 and 76 percent of the foods for which we have item prices, and retail items without environmental data include a variety of processed foods with important impacts along supply chains after leaving the farm. Future studies could conduct and assemble a wider range of assessments for more food items at each location, to allow more precise matching of environmental footprints with item prices observed at that place and time.

A second limitation is that item prices often refer to food items that can be made using various production processes with different environmental footprints. For example, fish prices represent an average over both farmed and wild-caught fish, but water footprint estimates were available only for farmed fishes. For most food items, average prices also combine locally produced and imported items produced using a variety of practices, including conventional, organic, or regenerative methods, as well as differing amounts of irrigation and energy use in greenhouses, transportation, and other stages. (Appendix 1c contains details on how item prices are matched to environmental footprint estimates.) Some food is also grown or processed at home, in contrast to the market purchases that are the focus of this study. Future work could focus on the prices, environmental footprints, and nutritional quality of foods produced using specific practices in different parts of the world.

Third, we use combined water footprint estimates, which do not differentiate between green, blue, and grey water use and do not account for local water scarcity. The comprehensive water footprint indicator was designed to represent total direct and indirect water use to produce a product over the entire supply chain (43), though separate analysis of green, blue, and grey water use may provide additional insights. At present, estimates of each type of water use for each food item and country at the system boundaries of this study are not yet available. Food systems also have environmental impacts beyond greenhouse gas emissions and water use, such as biodiversity loss and pollution, for which estimates of the magnitude of specific food items' impacts are sparse. As additional studies identify how specific foods in specific contexts contribute to these environmental impacts, our approach could be extended to these additional dimensions.

*Conclusions*

Reducing the environmental footprint and improving the nutritional profile of human diets is an important priority for many consumers, researchers, and policymakers. This study compares the retail food options sold around the world and finds that lower-priced items have systematically lower environmental footprints within almost all food groups, across the full range of item prices. In contrast, we find smaller and less consistent correlations between item prices and nutritional profile scores. More expensive alternatives in the same food group sometimes have higher nutritional profile scores; in other cases, more expensive options have lower or similar nutritional profiles scores as less expensive options. Taking account of these differences in environmental harm, health attributes, and price by type of food could help guide consumer choice, food businesses, and policy interventions towards healthier and more environmentally sustainable options for all.

**CRediT statement:** EMM, NTB, and WAM conceptualized and designed the study. EMM acquired and analyzed the data. All authors interpreted the results. EMM drafted the manuscript. All authors revised the manuscript for important intellectual content and read and approved the final manuscript as submitted. WAM obtained funding for the study. EMM and WAM have accessed and verified the data. All authors were responsible for the decision to submit the manuscript.

**Data availability:** ICP food prices (33) were obtained under a confidentiality agreement as specified here: https://www.worldbank.org/en/programs/icp/data#3. Food composition data are available online from the USDA (35), FAO (36), and national sources (37). Carbon and water footprints are available in the supplement of Petersson et al. (2021) (39,40). Food Compass Scores, Health Star Ratings, and Nutri-Scores are available in the supplement of Barrett et al. (2024) (43); Food Compass Scores by domain were provided by the author and are available upon request. Coding files for this analysis will be made available upon publication of the article, at https://sites.tufts.edu/foodpricesfornutrition and https://osf.io/zbqja.

**Funding:** This work was funded by the Gates Foundation and the UK government through the Innovative Methods and Metrics for Agriculture and Nutrition Actions project (INV-002962) and the Food Prices for Nutrition project (INV-016158), with additional support from the World Bank and the Food and Agriculture Organization of the United Nations.

**Conflicts of interest:** The authors declare no conflicts of interest.

**Declaration of generative AI and AI-assisted technologies in the writing process:** The authors did not use generative AI or AI-assisted technologies in the process of writing, reviewing, or editing this content.

**Table 1. Mean, standard deviation, minimum, and maximum of the price, carbon footprint, water footprint, and Food Compass Score of retail food items**

| | Food group | N | Mean | SD | Min | Max |
|---|---|---|---|---|---|---|
| Price per 100 kcal (2017 USD) | Starchy staples | 9758 | 0.20 | 0.17 | 0.01 | 2.09 |
| | Animal-source foods | 17126 | 1.35 | 1.84 | 0.03 | 37.60 |
| | Legumes, nuts & seeds | 1820 | 0.44 | 0.66 | 0.02 | 9.18 |
| | Vegetables | 5667 | 1.74 | 1.69 | 0.04 | 25.42 |
| | Fruits | 5136 | 1.05 | 0.82 | 0.02 | 9.04 |
| | Oils & fats | 2805 | 0.17 | 0.13 | 0.01 | 1.79 |
| | Sugars, sweets & candies | 6004 | 0.46 | 0.43 | 0.01 | 5.19 |
| Carbon footprint per 100 kcal (g CO2e) | Starchy staples | 9003 | 0.038 | 0.017 | 0.011 | 0.11 |
| | Animal-source foods | 13925 | 0.51 | 0.56 | 0.053 | 8.1 |
| | Legumes, nuts & seeds | 1440 | 0.032 | 0.053 | 0.012 | 0.30 |
| | Vegetables | 4746 | 0.16 | 0.13 | 0.032 | 0.94 |
| | Fruits | 4714 | 0.11 | 0.14 | 0.016 | 0.62 |
| | Oils & fats | 2687 | 0.051 | 0.041 | 0.011 | 0.12 |
| | Sugars, sweets & candies | 3705 | 0.066 | 0.063 | 0.020 | 0.21 |
| Water footprint per 100 kcal (L) | Starchy staples | 8893 | 44 | 16 | 21 | 160 |
| | Animal-source foods | 13906 | 320 | 260 | 39 | 1600 |
| | Legumes, nuts & seeds | 1379 | 180 | 76 | 47 | 380 |
| | Vegetables | 5190 | 120 | 73 | 23 | 1100 |
| | Fruits | 4709 | 170 | 110 | 47 | 470 |
| | Oils & fats | 2687 | 80 | 44 | 29 | 250 |
| | Sugars, sweets & candies | 2781 | 49 | 17 | 17 | 93 |
| Food Compass Score | Starchy staples | 8650 | 33 | 21 | 1 | 89 |
| | Animal-source foods | 16862 | 60 | 23 | 1 | 100 |
| | Legumes, nuts & seeds | 1812 | 90 | 8 | 40 | 100 |
| | Vegetables | 5171 | 93 | 13 | 41 | 100 |
| | Fruits | 5110 | 81 | 21 | 1 | 100 |
| | Oils & fats | 2638 | 54 | 26 | 13 | 78 |
| | Sugars, sweets & candies | 6004 | 14 | 19 | 1 | 91 |

*Note: Estimates were rounded to reflect the precision of the underlying datasets (2 decimal places for price, 2 significant figures for carbon and water footprint, 0 decimal places for Food Compass Score).*

**Figure 1. Estimated mean carbon footprint, water footprint, and nutrient profile conditional on price per kilocalorie, by food group**

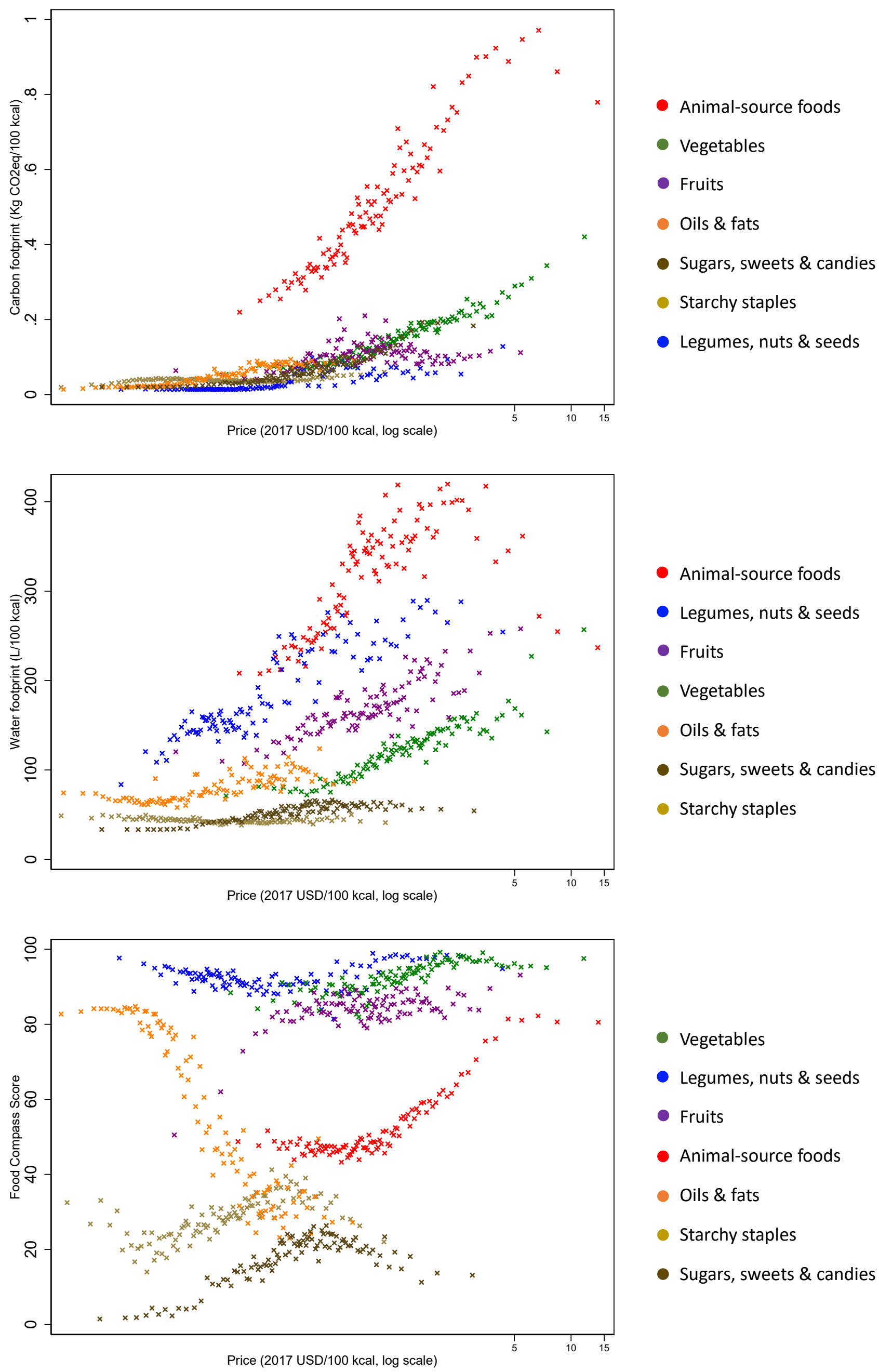

*Note: Carbon footprint and water footprint estimates from Petersson et al. (2021) and Food Compass Score calculations from Barrett et al. (2024) matched to average retail food prices from the World Bank International Comparison Program in 2011 and 2017 for 811 food items in 181 countries. Price in 2017 USD per kilocalorie is shown in natural-log scale. Figures are binned scatter plots, where each food group is represented by 100 data points, each of which is the mean value of the y-axis variable at the mean level of price per 100kcal across 100 equal-sized bins of price per 100kcal.*

**Figure 2. Estimated mean carbon footprint, water footprint, and nutrient profile conditional on price per recommended daily intake, by food group**

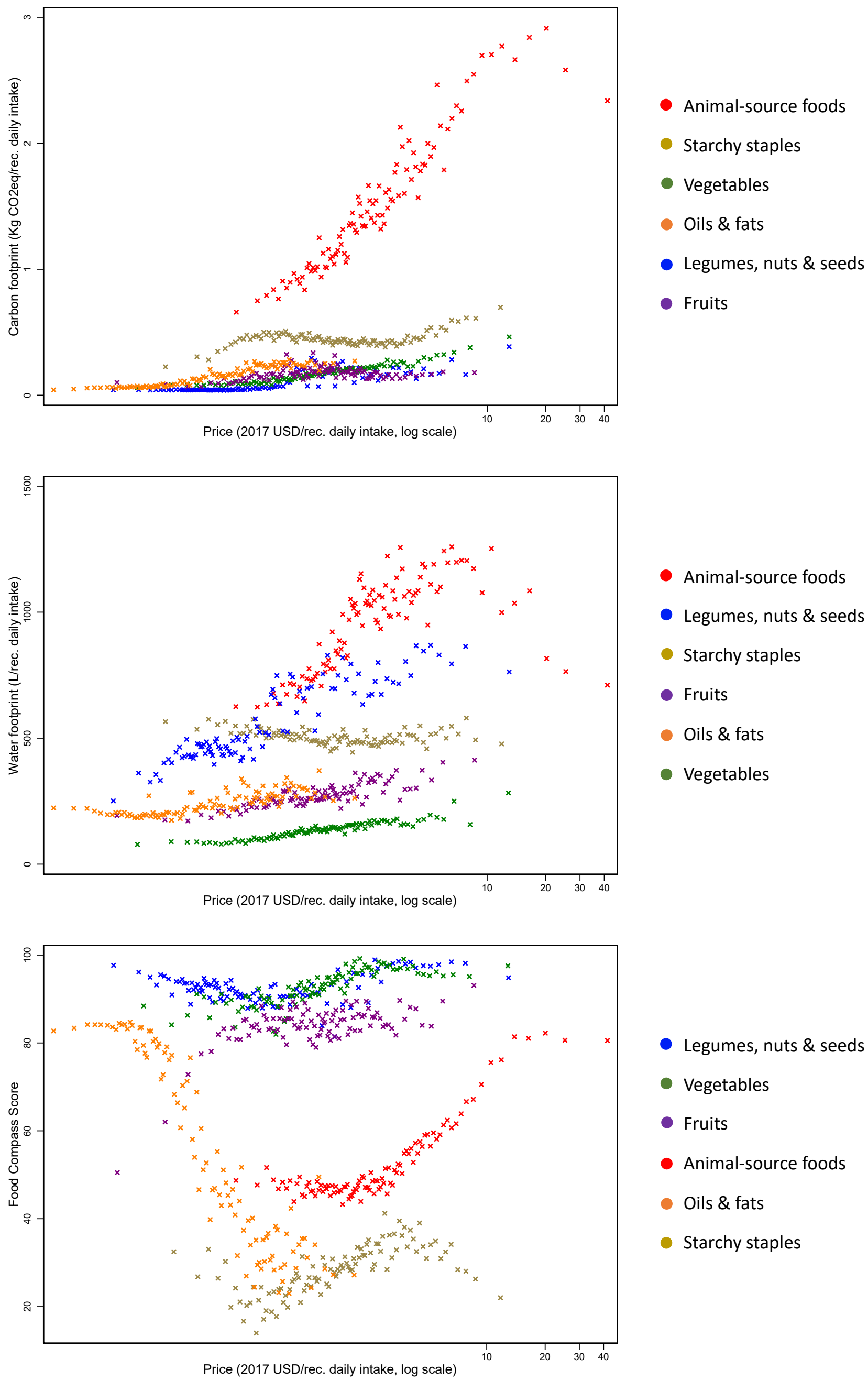

*Note: Carbon footprint and water footprint estimates from Petersson et al. (2021) and Food Compass Score calculations from Barrett et al. (2024) matched to average retail food prices from the World Bank International Comparison Program in 2011 and 2017 for 707 food items in 181 countries. Price in 2017 USD per recommended daily intake is shown in natural-log scale. Estimates per recommended daily intake omit "sugars, sweets & candies" because there is no recommended intake of this food group. Figures are binned scatter plots, where each food group is represented by 100 data points, each of which is the mean value of the y-axis variable at the mean level of price per recommended daily intake across 100 equal-sized bins of price per recommended daily intake.*

**Figure 3. Associations between price per kilocalorie and carbon footprint, water footprint, and Food Compass Score by food group**

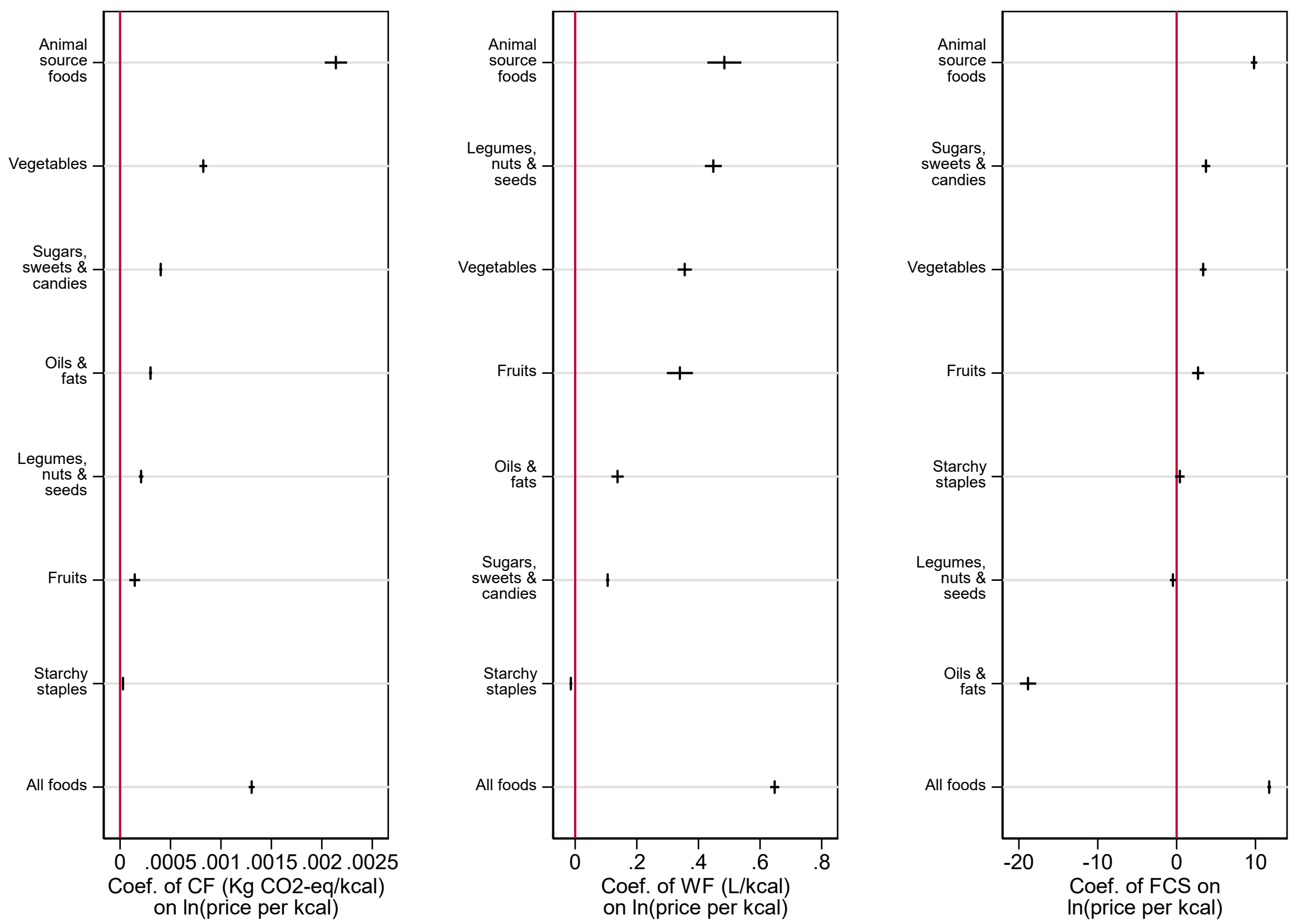

*Note: Tick marks represent coefficients and 95% confidence intervals of linear regressions of carbon footprint (CF), water footprint (WF), and Food Compass Score (FCS) on log(price) with country fixed effects, stratified by food group and for all foods.*

**Figure 4. Associations between the 9 domains of Food Compass Score and price per kilocalorie by food group**

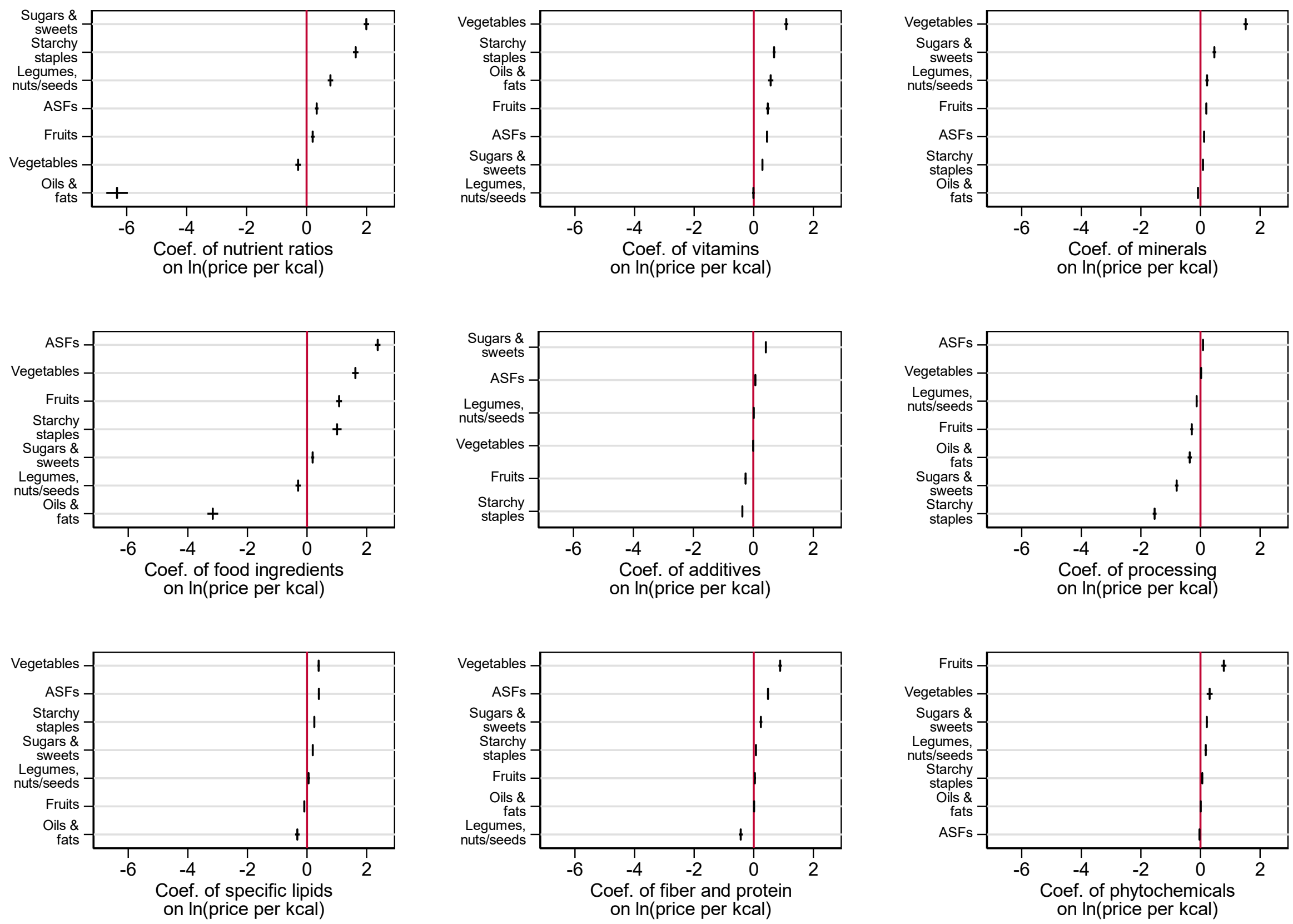

*Note: Tick marks represent coefficients and 95% confidence intervals of linear regressions of scores in each domain of Food Compass Score (FCS) on log(price) with country fixed effects, stratified by food group. Descriptions of the 9 domains of FCS are available in Appendix 1d and scoring details are in Mozaffarian et al. 2021. Food compass domains include (1) favorable nutrient ratios, including unsaturated:saturated fats, carbohydrates:fiber; potassium:sodium; (2) vitamins related to undernutrition and chronic disease; (3) minerals related to undernutrition and chronic disease; (4) food-based ingredients with impacts on chronic diseases; (5) additives with evidence of health harms; (6) processing characteristics with health implications; (7) specific lipids with health associations; (8) total fiber and protein; (9) phytochemicals, including flavonoids and carotenoids. Estimates for coefficients on "additives" omit the "oils & fats" food group and estimates for coefficients on "specific lipids" omit the "legumes, nuts and seeds," "vegetables," and "fruits" food groups because all food items scored zero in these dimensions. ASF stands for "animal-source foods."*

# Supplemental materials

**Retail prices, environmental footprints, and nutritional profiles of commonly sold retail food items in 181 countries**

Elena M. Martinez, PhD [1], Nicole Tichenor Blackstone, PhD [1 2], Parke E. Wilde, PhD [1], Anna W. Herforth, PhD [3], William A. Masters, PhD, FASN [1 4]

[1] Friedman School of Nutrition Science and Policy, Tufts University, USA
[2] Department of Biomedical Engineering, Tufts University, USA
[3] Food Prices for Nutrition Project, Boston, USA
[4] Department of Economics, Tufts University, USA

Correspondence: Elena M. Martinez, Elena.Martinez@tufts.edu

**Table of Contents**

## Appendix 1. Supplementary methodological details

### Appendix 1a. Equations for calculating price per kilogram (kg), kilocalorie (kcal), and recommended daily intake

We use the following equations to convert the price of each food item from the price per kilogram of the food item to price per price per kilocalorie and price per recommended daily intake.

$$Price\ per\ kg\ of\ edible\ matter = \frac{Price\ per\ kg\ of\ food}{Edible\ portion\ of\ food}$$

For each food item, we divide the price per kilogram of food by the edible portion of that food item to estimate the price per kilogram of the portion of the food item that is considered edible. The edible portion of a food item is the portion that is considered edible. Food composition table report the percentage of each food item considered refuse; edible portion is 100% minus the refuse percentage.

$$Price\ per\ kcal\ of\ edible\ matter = \frac{Price\ per\ kg\ of\ edible\ matter}{Kcal\ per\ kg\ of\ food}$$

For each food item, we divide the price per kilogram of edible matter by the kilocalories per kilogram of that food item to estimate the price per kilocalorie of the portion of the food item that is considered edible.

$$Price\ per\ recommended\ daily\ intake = (Price\ per\ kcal\ of\ edible\ matter) \times (Recommended\ intake\ in\ kcal\ of\ food\ group)$$

For each food item, we multiply the price per kilocalorie by the recommended intake in kilocalories for each food group to estimate the price per recommended daily intake of the food item. Recommended intakes are based on the Healthy Diet Basket (Herforth et al. 2022).

The edible portion and kilocalories per kilogram of each food item were retrieved from the USDA National Nutrient Database for Standard Reference Release 28 (SR-28), the FAO/INFOODS Food Composition Table for Western Africa, the Food Composition Table for Bangladesh, the FAO/INFOODS Global food composition database for fish and shellfish (uFish 1.0), the USDA Food and Nutrient Database for Dietary Studies (FNDDS) 2019-20, and USDA Food Data Central. This methodology for matching ICP food items to food composition data is based on Appendix 3 of Hirvonen et al. (2020).

**Appendix 1b. Healthy diet basket daily recommended intakes by food group**

| Food group | Minimum number of food items selected for cost of healthy diet | Total energy content (kcal) | Equivalent gram content, by reference food (edible portion) |
|---|---|---|---|
| Starchy staples | 2 | 1160 | 322g dry rice |
| Animal-source foods | 2 | 300 | 210g egg |
| Legumes, nuts, and seeds | 1 | 300 | 85g dry bean |
| Vegetables | 3 | 110 | 270-400g vegetable |
| Fruits | 2 | 160 | 230-300g fruit |
| Oils and fats | 1 | 300 | 34g oil |

Source: Herforth et al., 2022

## Appendix 1c: Environmental impact data sources and matching

ICP food items were matched to food item names in Petersson et al. (2021). Where possible, ICP names were matched directly to names used by Petersson. If a direct match to the food item was not available, we matched to estimates of GHG emissions and water footprint for a group of foods (e.g., berries, seafood), referred to as typology or sub-typology by Petersson et al. (For example, a food item "raspberries" might fit in the typology "fruits" and the subtypology "berries.") For example, shrimp and prawns were matched directly to an estimate of GHG emissions for shrimp and prawns, while crab was matched to an estimate of GHG emission for seafood on average. ICP food items were excluded from the analysis if there was no relevant food item, typology, or subtypology in Petersson et al. (e.g., camel meat) or if the relevant typology or subtypology did not account for important ingredients or value chain stages. For example, dried fish, smoked fish, and canned fish other than tuna were excluded because the Petersson et al. estimate of GHG emissions for processed fish included only estimates for canned tuna and fish sticks.

Petersson et al. included estimates of the certainty of each GHG emissions and water footprint estimate, along with suggestions for whether to use the estimate at the item, typology, or subtypology level. We followed the following rules to match food item, typology, and subtypology estimates to each food item.

| **Recommendation in Petersson et al. (2021) database** | **Estimate used** |
|---|---|
| "Ok item" | Food item |
| "Item matched typology" OR "Better typology" | Typology |
| "Better subtypology" or "Better typology or subtypology" | Subtypology |
| "Item or typology" or "Item or typology or subtypology" | Food item, if item estimate had low uncertainty;<br>Typology or subtypology, if item estimate had high uncertainty |

## Appendix 1d. Food Compass Score, Nutri-Score, and Health Star Rating

We estimate the nutritional profile of food items using 3 established metrics: Food Compass Score, Nutri-Score, and Health Star Rating.

Food Compass Score (FCS) is a nutrient profiling system that rates the healthfulness of foods on a scale of 0-100 based on 9 domains relevant to health outcomes, including nutrient ratios, vitamins, minerals, food-based ingredients, additives, processing, specific lipids, total fiber and protein, and phytochemicals (Mozaffarian et al. 2021; Barrett et al. 2024). The 9 Food Compass Score domains are shown below. Each domain is composed of a set of attributes; the full list of attributes for each domain is available in Mozaffarian et al. (2021), Supplementary Table 3. For a given food item, each attribute is scored on a scale from -10 to 10. The scoring algorithm and potential ranges for each attribute are described in Mozaffarian et al. (2021), Supplementary Table 3. The score for each domain is calculated as the average score of the attributes in that domain. The 9 domain scores are summed to provide the FCS for each food item. The final FCS is then scaled from 0 to 100 for ease of interpretation. Minor updates to the FCS domains are described in Barrett et al. (2024), Supplementary Table 1.

| **Domain** | **Description of attributes** |
|---|---|
| Nutrient ratios | Ratios of the quality of fats (unsaturated:saturated fats), carbohydrates (carbohydrate:fibre), and/or minerals (potassium:sodium) |
| Vitamins | Vitamins related to undernutrition and chronic diseases (e.g., Vitamin A, thiamin) |
| Minerals | Minerals related to undernutrition and chronic diseases (e.g., calcium, iron) |
| Food-based ingredients | Food groups with impacts on chronic diseases (e.g., fruits, whole grains, refined carbohydrates) |
| Additives | Food additives with evidence of heath harms (e.g., nitrates, artificial sweeteners) |
| Processing | NOVA classification and other processing characteristics (e.g., fermentation, frying) with health implications |
| Specific lipids | Lipids with evidence of health associations (e.g., trans fats, cholesterol) |
| Fiber and protein | Total fiber and total protein |
| Phytochemicals | Total flavonoids and total carotenoids |

*Source: Mozaffarian et al. (2021), adapted from Supplementary Table 3*

Nutri-Score, created by Santé Publique France, is a nutritional rating from 0-5 based on the food item's content per 100g of nutrients and foods to promote, including dietary fiber, protein, fruits, vegetables, pulses, nuts, and plant oils, and nutrients to limit, including total sugar, saturated fat, sodium, and total energy. The Nutri-Score is translated into a letter from A to E for use on a color-coded front-of-pack label (Santé Publique France, 2023).

Health Star Rating is a nutritional rating that scores foods between 0.5 and 5 to inform front-of-pack food labels with 0.5 to 5 stars. Health Star Ratings are based on the food item's total energy; content of nutrients associated with chronic disease, including saturated fat, sodium, and sugar; and content of nutrients and foods associated with improved health outcomes, including fiber, protein, fruits, vegetables, nuts, and legumes (Australian Government, 2023).

**Appendix 1e. Data inclusion flow chart**

We analyze prices for retail food and beverage items from 175 countries in 2011 and 177 countries in 2017 from the International Comparison Program (ICP), matched to food composition, carbon footprint (CF), water footprint (WF), and Food Compass Score (FCS) estimates. This flow chart shows how food price observations were selected for inclusion in the analysis.

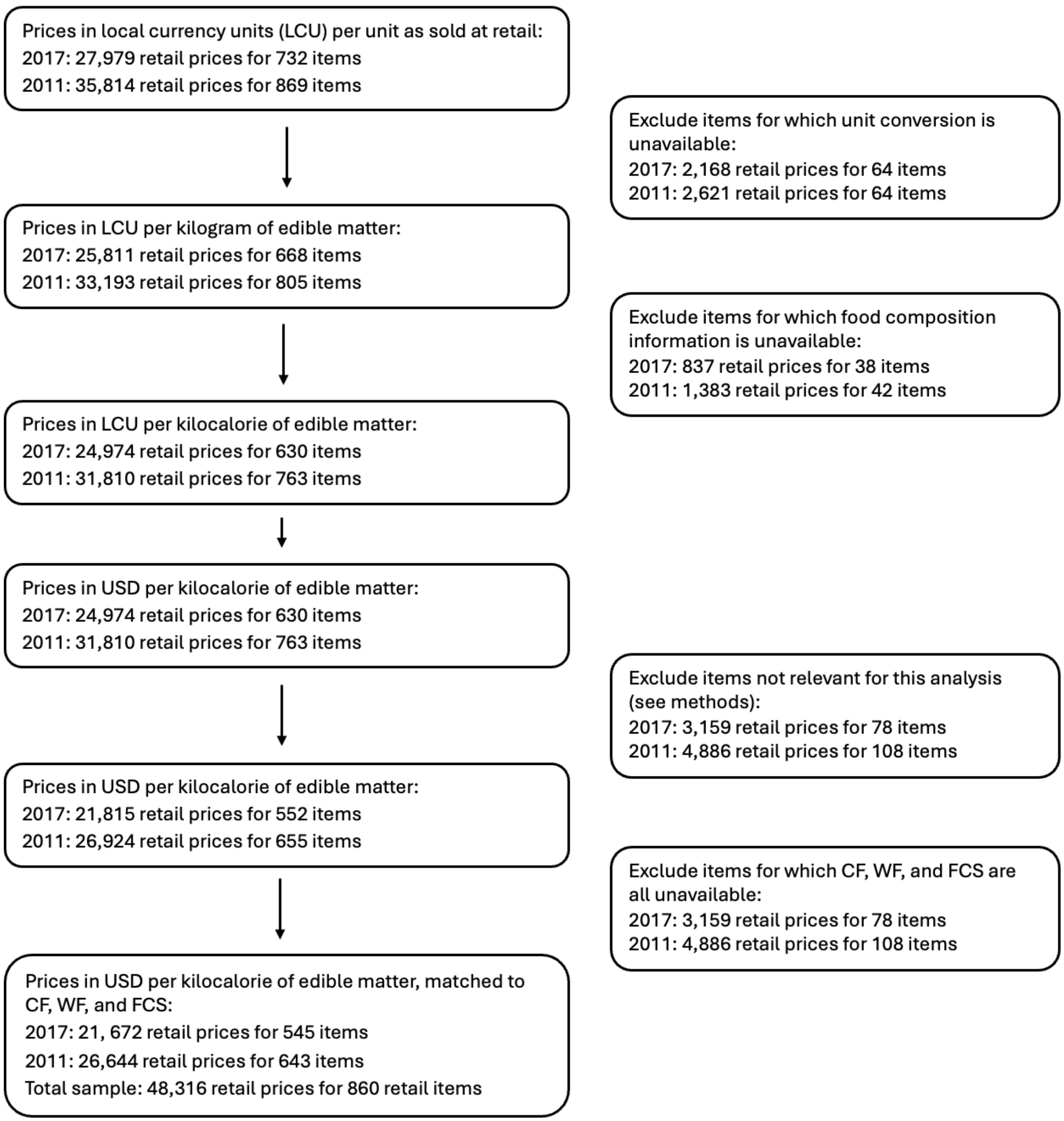

## Appendix 2: Alternative visualizations for comparing price, carbon footprint, water footprint, and Food Compass Score by food group

### Supplementary Figure 2a. Estimated mean carbon footprint and water footprint conditional on Food Compass Score, by food group

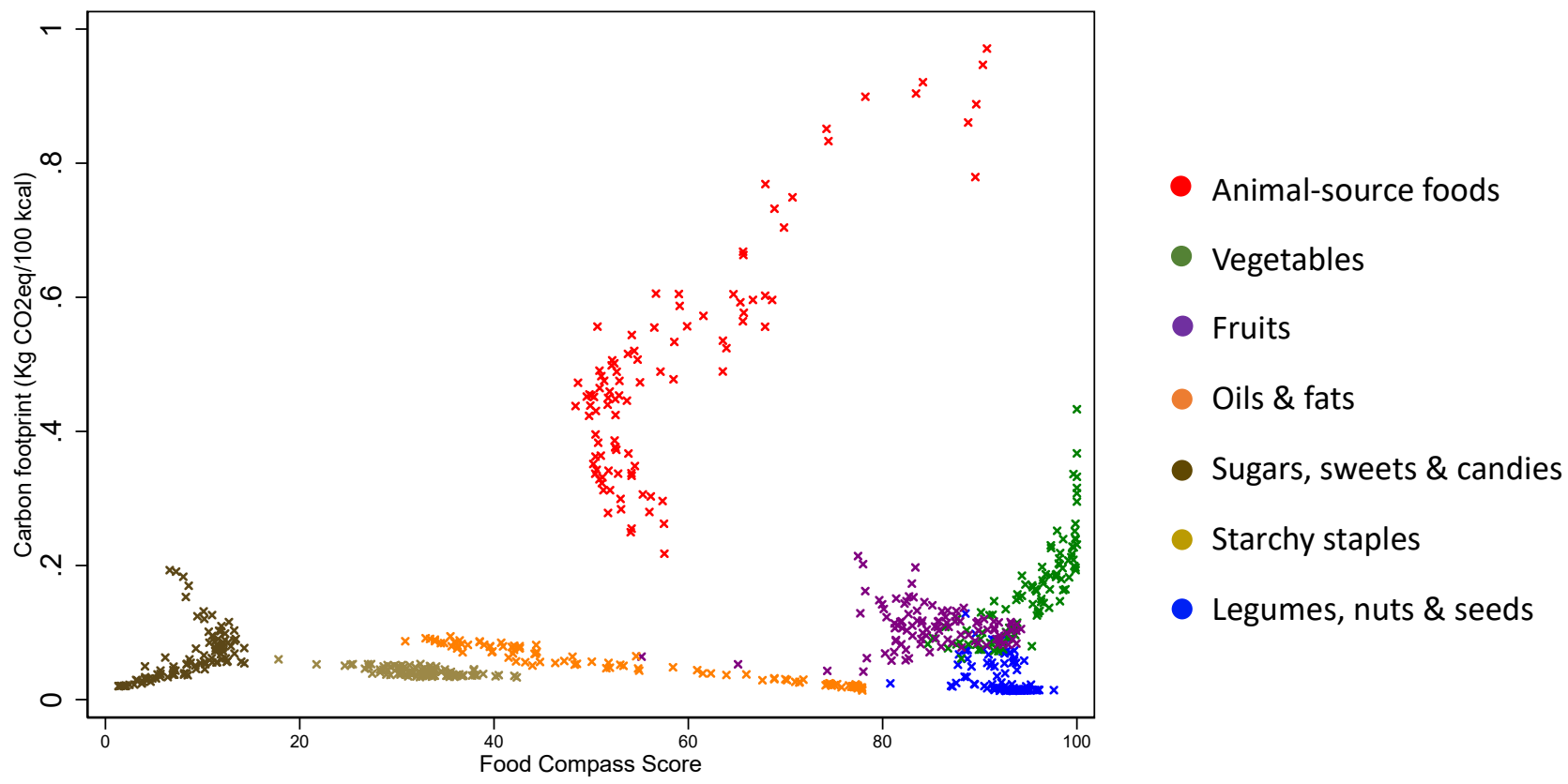

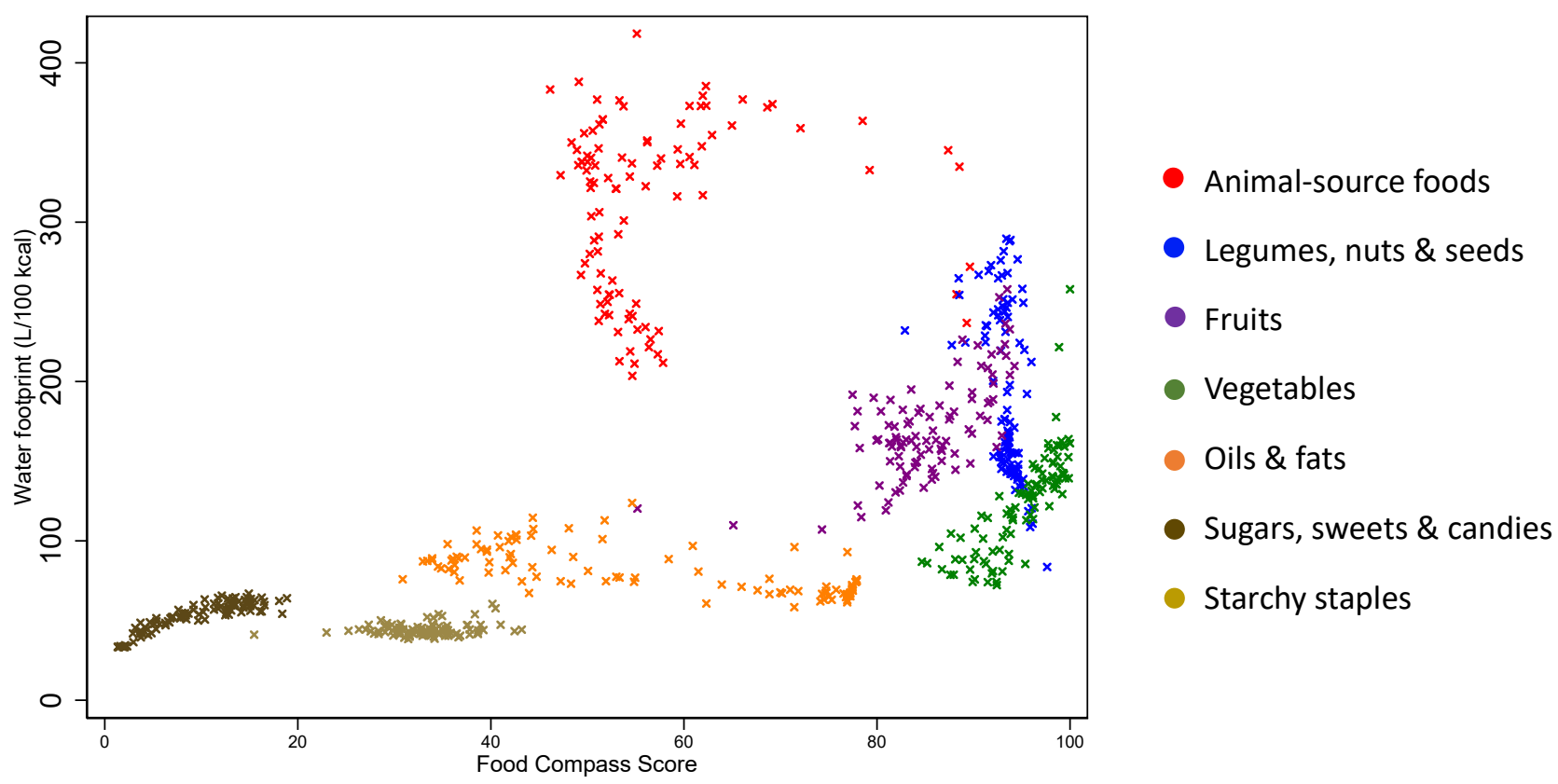

*Note: Carbon footprint and water footprint estimates from Petersson et al. (2021) and Food Compass Score calculations from Barrett et al. (2024) for 652 food items in 181 countries. Figures are binned scatter plots, where each food group is represented by 100 data points, each of which is the mean value of the y-axis variable at the mean level of FCS across 100 equal-sized bins of FCS.*

**Supplementary Figure 2b. Carbon footprint, water footprint, and Food Compass Score by deciles of price per kilocalorie and food group**

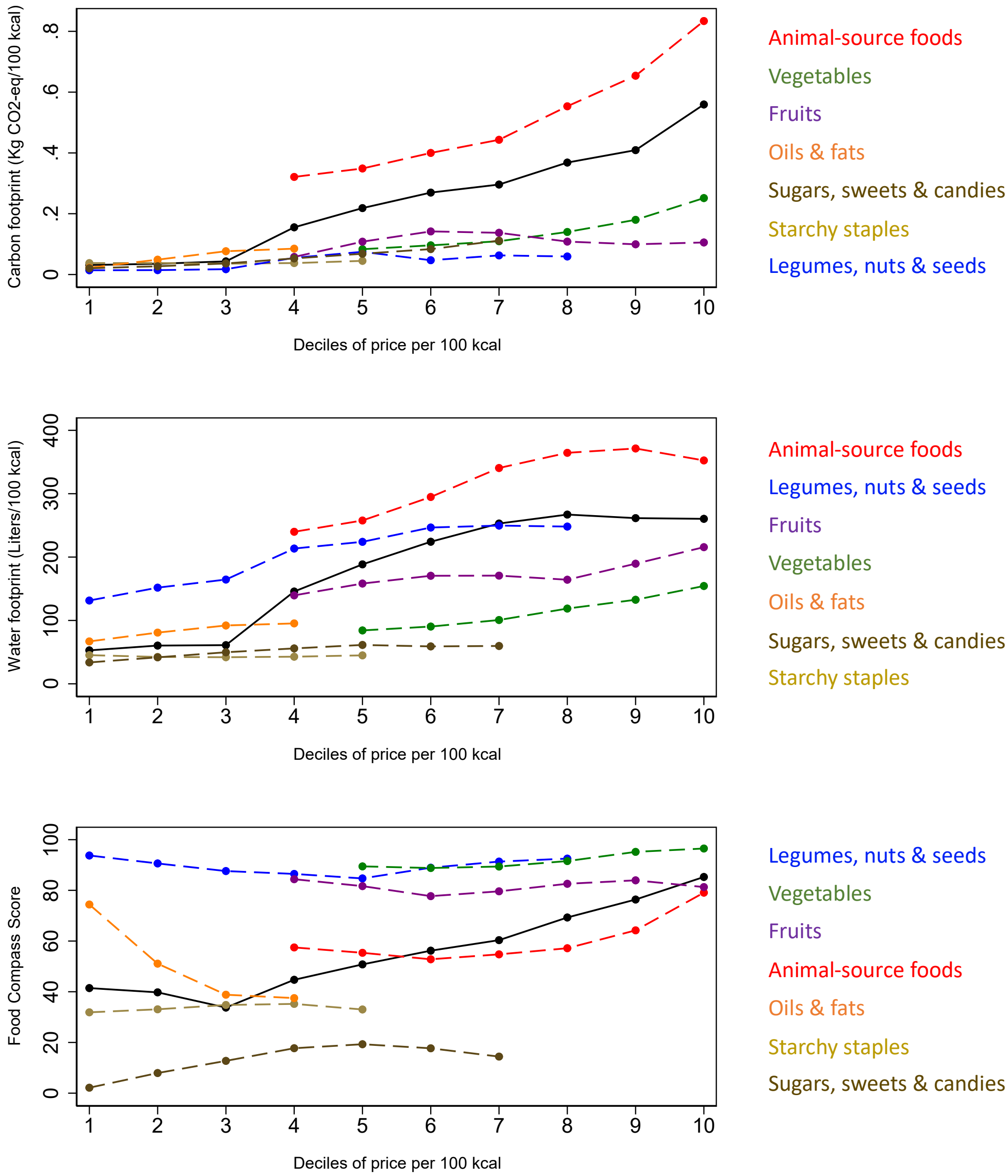

*Note: Carbon footprint and water footprint estimates from Petersson et al. (2021) and Food Compass Score calculations from Barrett et al. (2024) matched to average retail food prices from the World Bank International Comparison Program in 2011 and 2017 for 799 food items in 181 countries. Deciles represent deciles of price per 100 kcal by country and year; deciles that contain less than 5 percent of the observations for a food group are omitted.*

## Appendix 3: Comparing retail food prices, carbon footprint, and water footprint, and nutrient profile per kilogram and per recommended daily intake

**Supplementary Figure 3a. Estimated mean carbon footprint, water footprint, and nutrient profile conditional on price per kilogram, by food group**

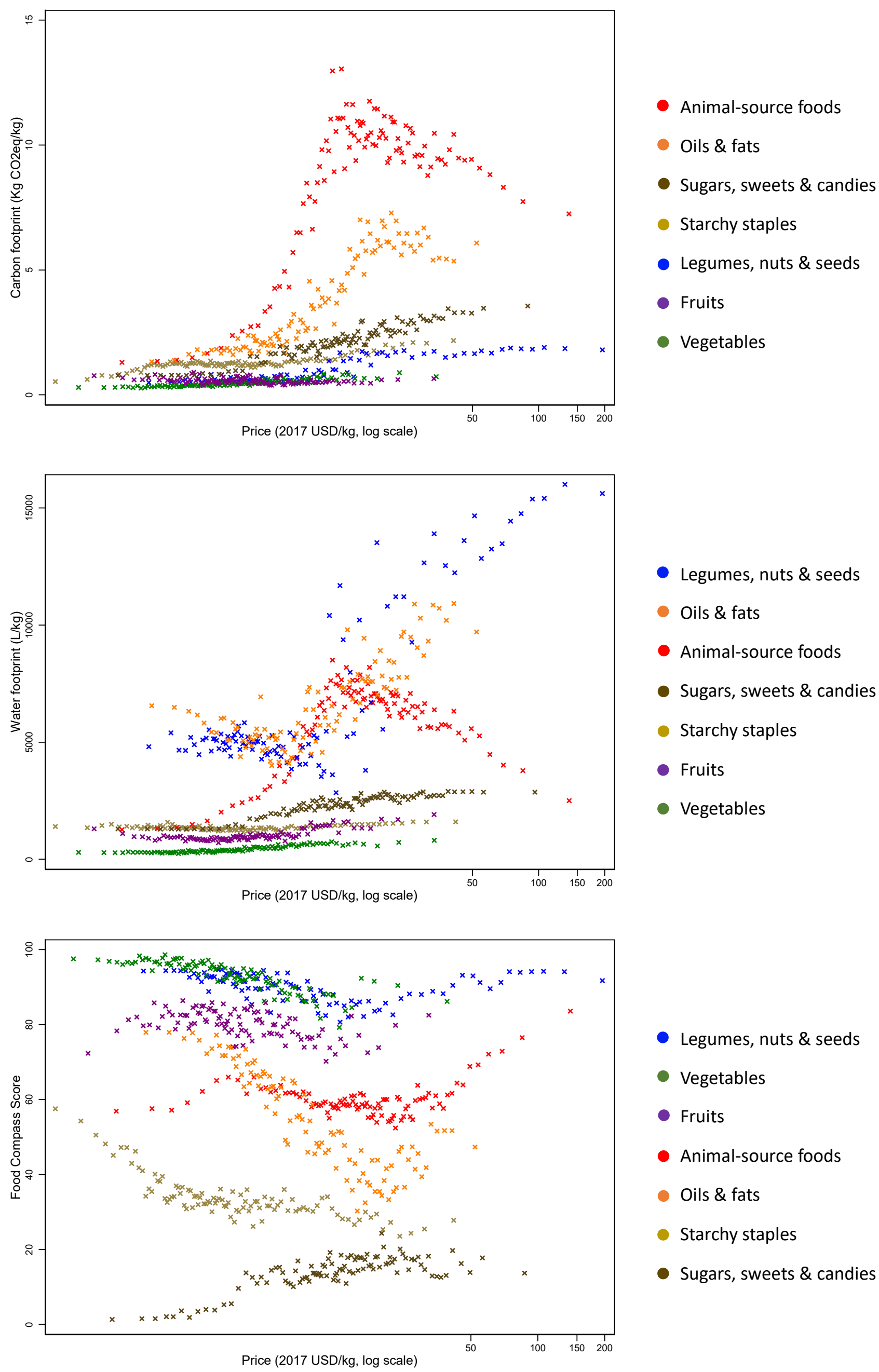

*Note: Carbon footprint and water footprint estimates from Petersson et al. (2021) and Food Compass Score calculations from Barrett et al. (2024) matched to average retail food prices from the World Bank International Comparison Program in 2011 and 2017 for 818 food items in 181 countries. Price in 2017 USD per recommended daily intake is shown in natural-log scale. Figures are binned scatter plots, where each food group is represented by*

*100 data points, each of which is the mean value of the y-axis variable at the mean level of price per kilogram across 100 equal-sized bins of price per kilogram.*

**Supplementary Figure 3b. Estimated mean carbon footprint and water footprint conditional on Food Compass Score per kilogram, by food group**

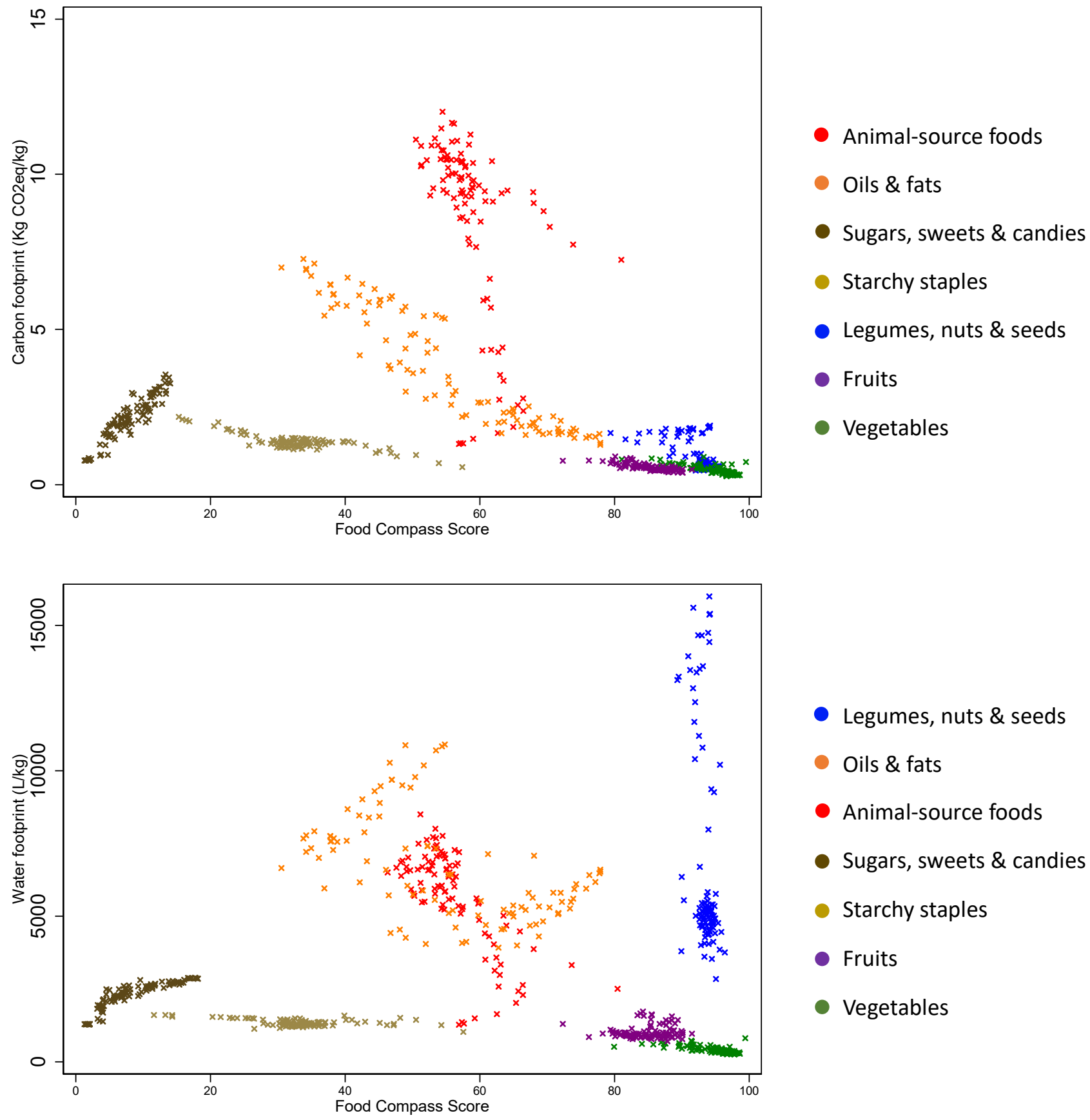

*Note: Carbon footprint and water footprint estimates from Petersson et al. (2021) and Food Compass Score calculations from Barrett et al. (2024) for 654 food items in 181 countries. Figures are binned scatter plots, where each food group is represented by 100 data points, each of which is the mean value of the y-axis variable at the mean level of FCS across 100 equal-sized bins of FCS.*

**Supplementary Figure 3c. Estimated mean carbon footprint and water footprint conditional on Food Compass Score per recommended daily intake, by food group**

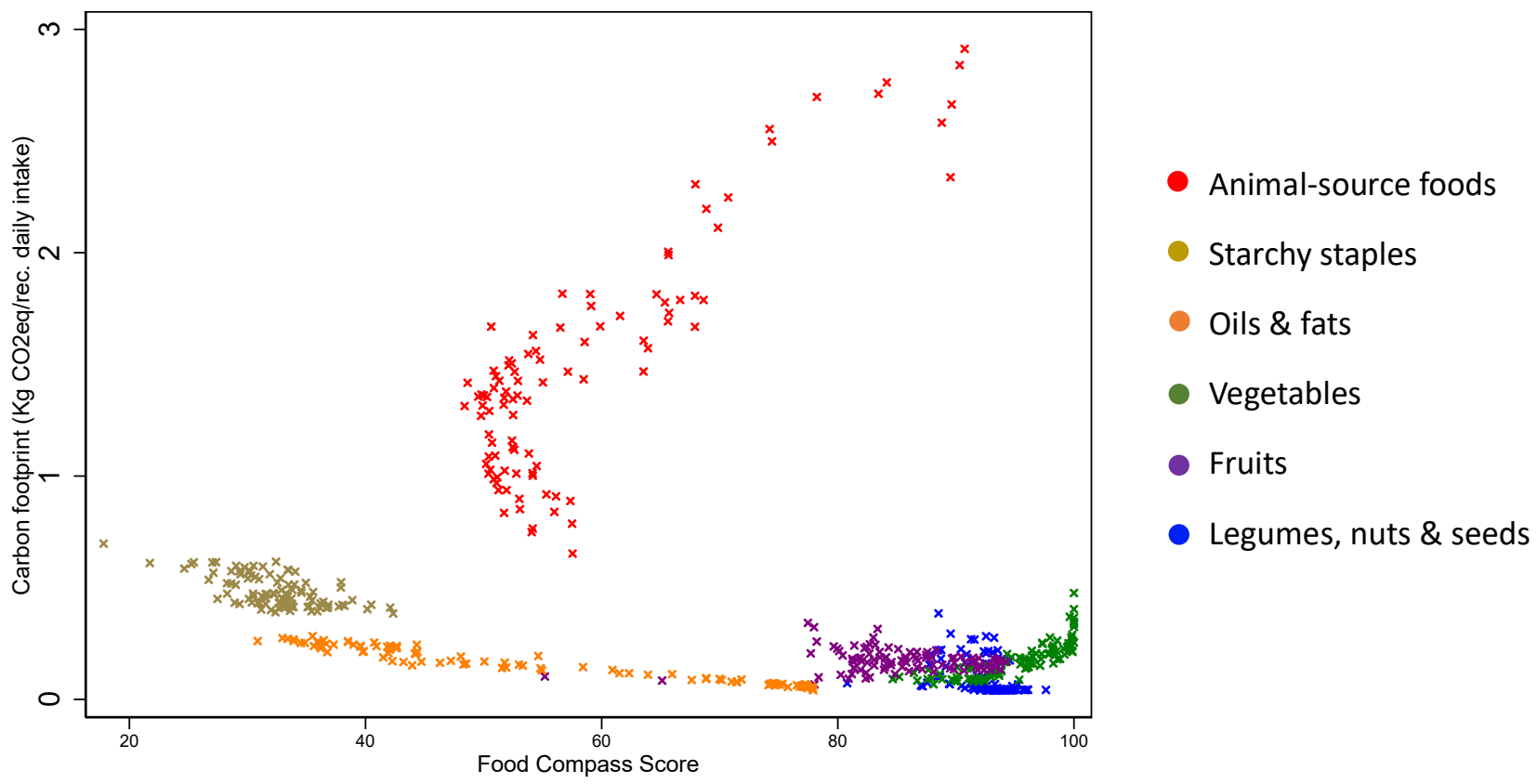

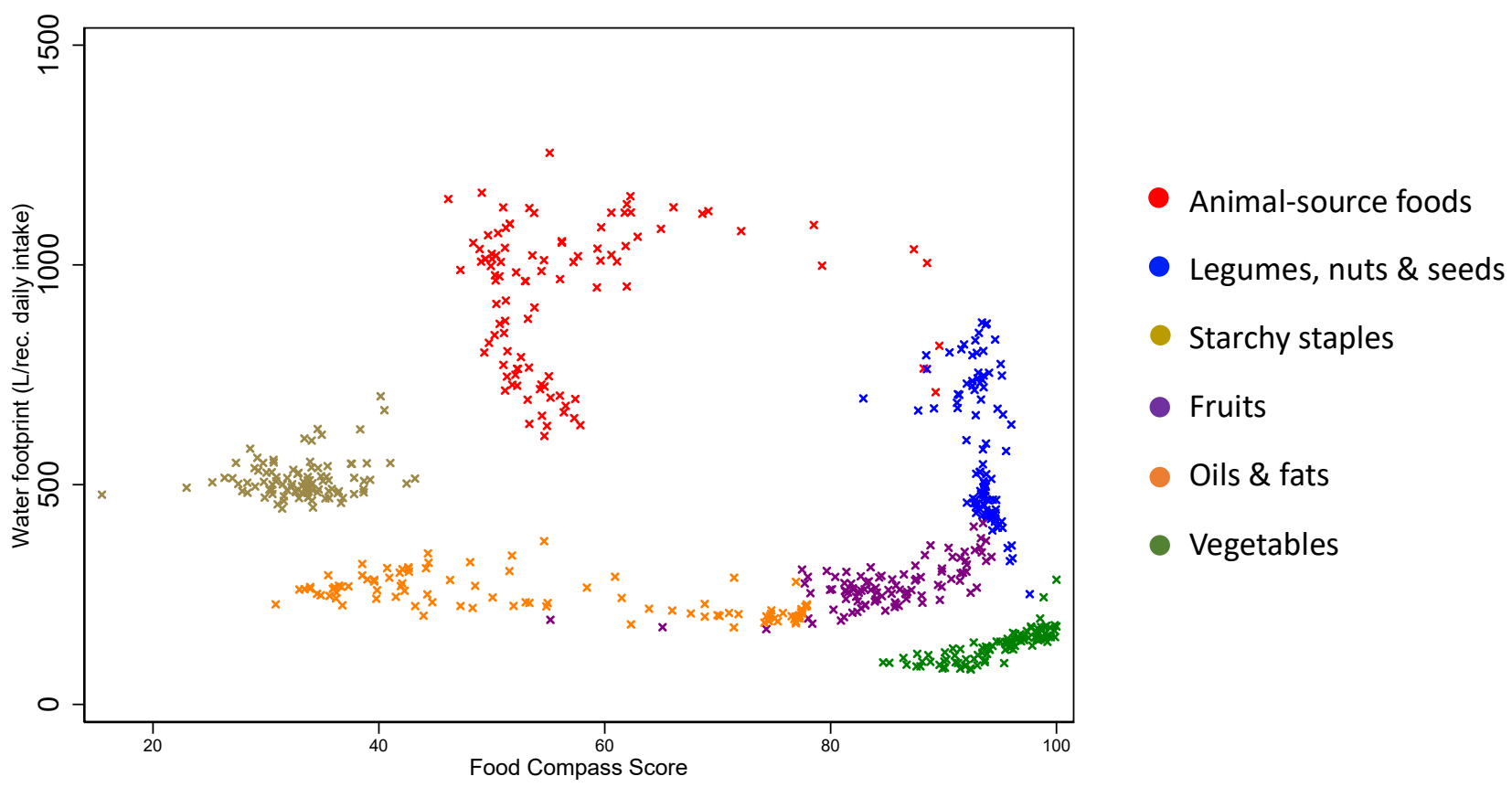

*Note: Carbon footprint and water footprint estimates from Petersson et al. (2021) and Food Compass Score calculations from Barrett et al. (2024) for 589 food items in 181 countries. Estimates per recommended daily intake omit "sugars, sweets & candies" because there is no recommended intake of this food group. Figures are binned scatter plots, where each food group is represented by 100 data points, each of which is the mean value of the y-axis variable at the mean level of FCS across 100 equal-sized bins of FCS.*

**Supplementary Figure 3d. Associations between price per kilogram and carbon footprint, water footprint, and Food Compass Score by food group**

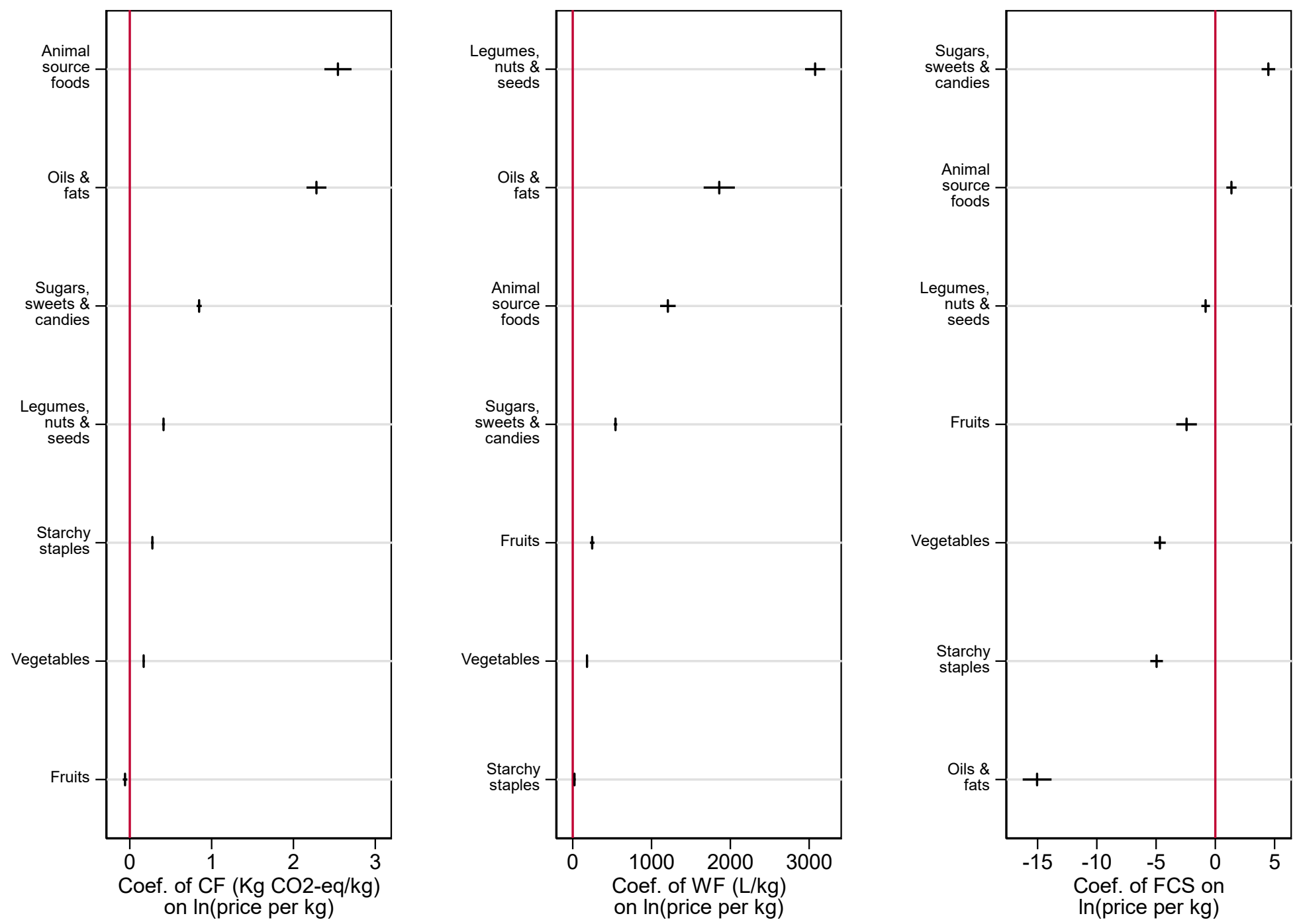

*Note: Tick marks represent coefficients and 95% confidence intervals of linear regressions of carbon footprint (CF), water footprint (WF), and Food Compass Score (FCS) on log(price) with country fixed effects, stratified by food group.*

**Supplementary Figure 3e. Associations between price per recommended daily intake and carbon footprint, water footprint, and Food Compass Score by food group**

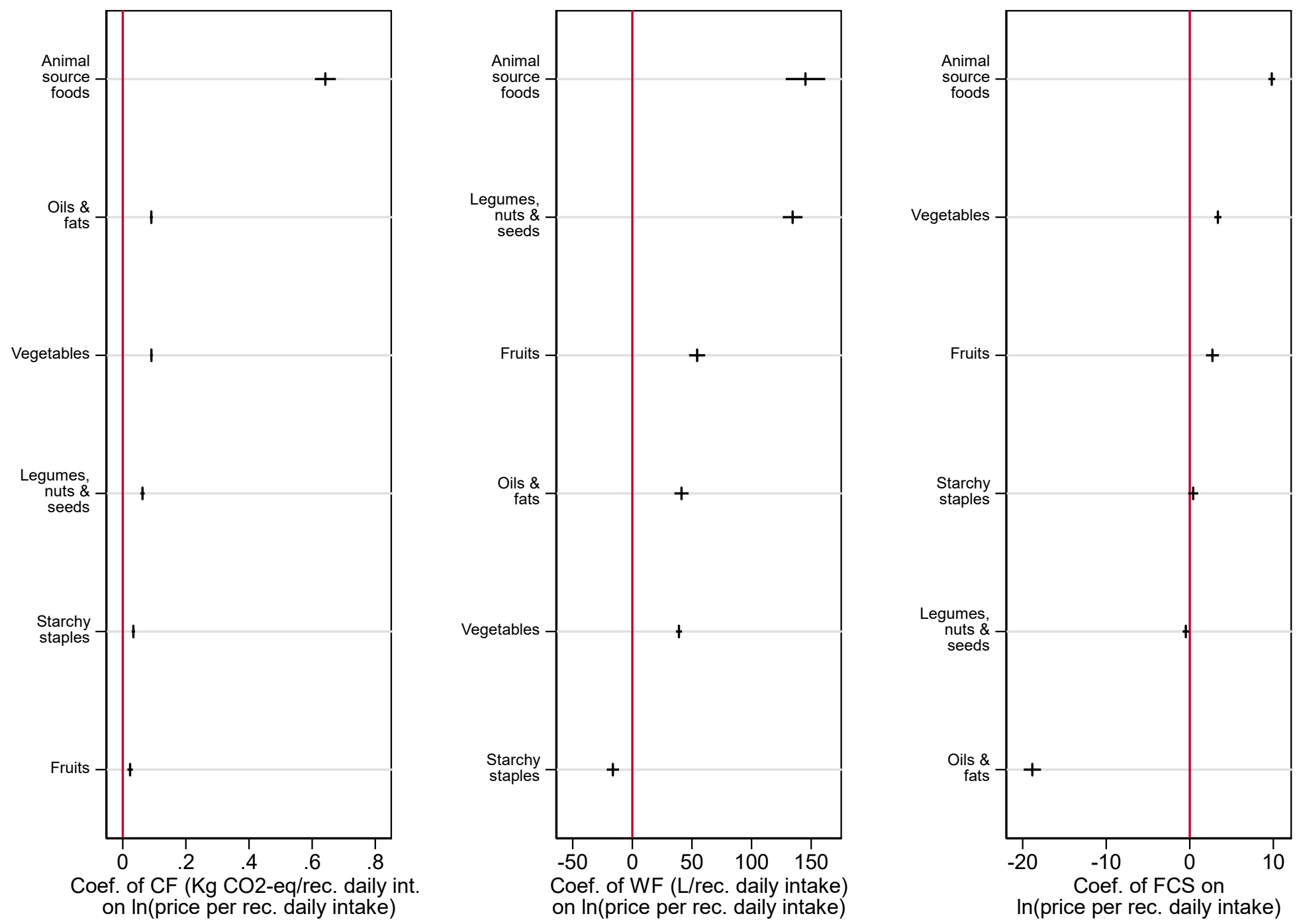

*Note: Tick marks represent coefficients and 95% confidence intervals of linear regressions of carbon footprint (CF), water footprint (WF), and Food Compass Score (FCS) on log(price) with country fixed effects, stratified by food group. Estimates per recommended daily intake omit "sugars, sweets & candies" because there is no recommended intake of this food group. Note that the between-group comparisons by recommended daily intake the same as between group-comparisons of associations between price per kilocalorie and CF, WF, and FCS because HDB recommended intakes are quantified in kilocalories per food group.*

**Supplementary Figure 3f. Carbon footprint, water footprint, and Food Compass Score by deciles of price per kilogram and food group**

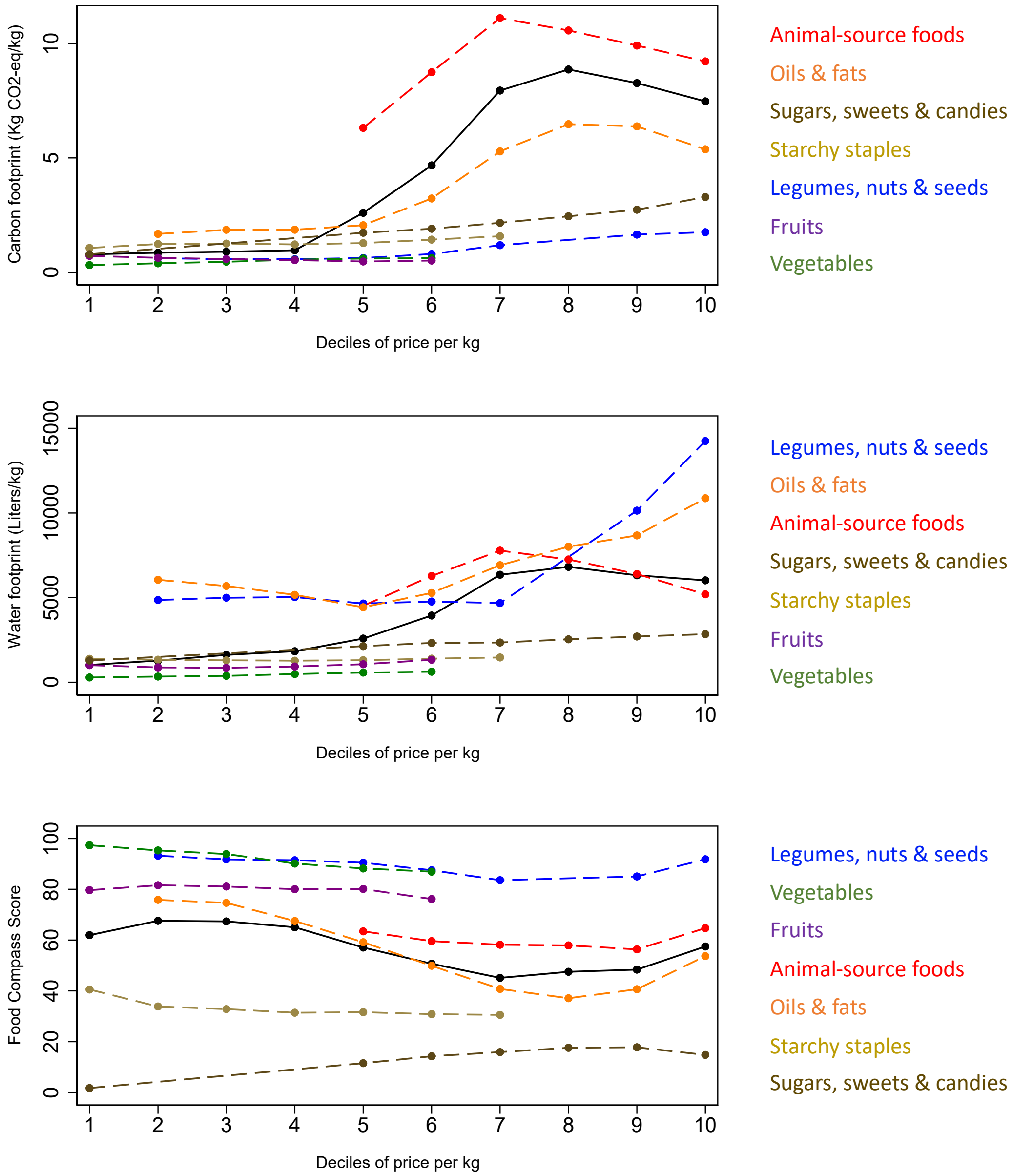

*Note: GHG emissions and water footprint estimates from Petersson et al. (2021) and Food Compass Score calculations from Barrett et al. (2024) matched to average retail food prices from the World Bank International Comparison Program in 2011 and 2017 for 793 food items in 181 countries. Deciles that contain less than 5 percent of the observations for a food group are omitted.*

**Supplementary Figure 3g. GHG emissions, water footprint, and Food Compass Score by deciles of price per recommended daily intake and food group**

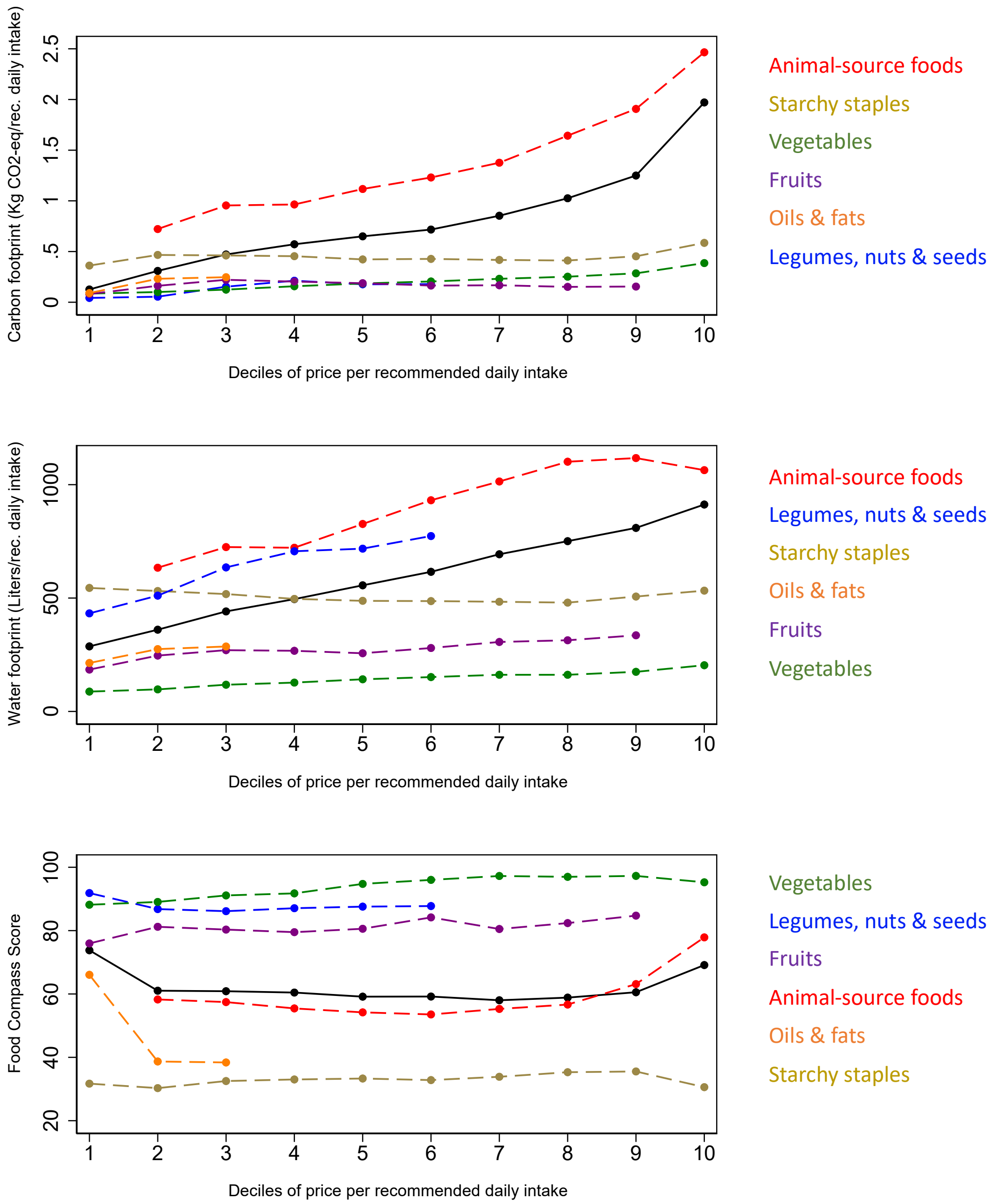

*Note: GHG emissions estimates from Petersson et al. (2021) matched to average retail food prices from the World Bank International Comparison Program in 2011 and 2017 for 698 food items in 181 countries. Estimates per recommended daily intake omit "sugars, sweets & candies" because there is no recommended intake of this food group. Deciles that contain less than 5 percent of the observations for a food group are omitted.*

## Appendix 4: Comparing retail food prices to nutritional profile using Health Star Rating and Nutri-Score

**Supplementary Table 4a. Mean, standard deviation, minimum, and maximum of the Food Compass Score, Health Star Rating, and NutriScore of retail food items**

| | | Starchy staples | Animal-source foods | Legumes, nuts & seeds | Vegetables | Fruits | Oils & fats | Sugars, sweets & candies |
|---|---|---|---|---|---|---|---|---|
| | N | 9758 | 17126 | 1820 | 5667 | 5136 | 2805 | 6004 |
| Food Compass Score | Mean | 33 | 60 | 90 | 93 | 81 | 54 | 14 |
| | SD | 21 | 23 | 8 | 13 | 21 | 26 | 19 |
| | Min | 1 | 1 | 40 | 41 | 1 | 13 | 1 |
| | Max | 89 | 100 | 100 | 100 | 100 | 78 | 91 |
| Health Star Rating | Mean | 3.4 | 3.3 | 4.8 | 4.6 | 4.0 | 2.3 | 1.4 |
| | SD | 0.7 | 1.3 | 0.5 | 0.3 | 0.7 | 1.2 | 0.8 |
| | Min | 1.0 | 0.5 | 1.5 | 3.5 | 0.5 | 0.5 | 0.5 |
| | Max | 5.0 | 5.0 | 5.0 | 5.0 | 5.0 | 3.5 | 3.5 |
| NutriScore | Mean | 4.0 | 3.2 | 4.5 | 4.9 | 4.2 | 1.7 | 1.7 |
| | SD | 0.9 | 1.2 | 0.9 | 0.4 | 1.2 | 0.5 | 0.6 |
| | Min | 1.0 | 1.0 | 2.0 | 3.0 | 1.0 | 1.0 | 1.0 |
| | Max | 5.0 | 5.0 | 5.0 | 5.0 | 5.0 | 2.0 | 4.0 |

*Note: Estimates were rounded to reflect the precision of the underlying datasets (0 decimal places for Food Compass Score, 1 decimal place for Health Star Rating and NutriScore).*

**Supplementary Figure 4a. Estimated mean Food Compass Score, Health Star Rating, and Nutri-Score conditional on price per kilocalorie, by food group**

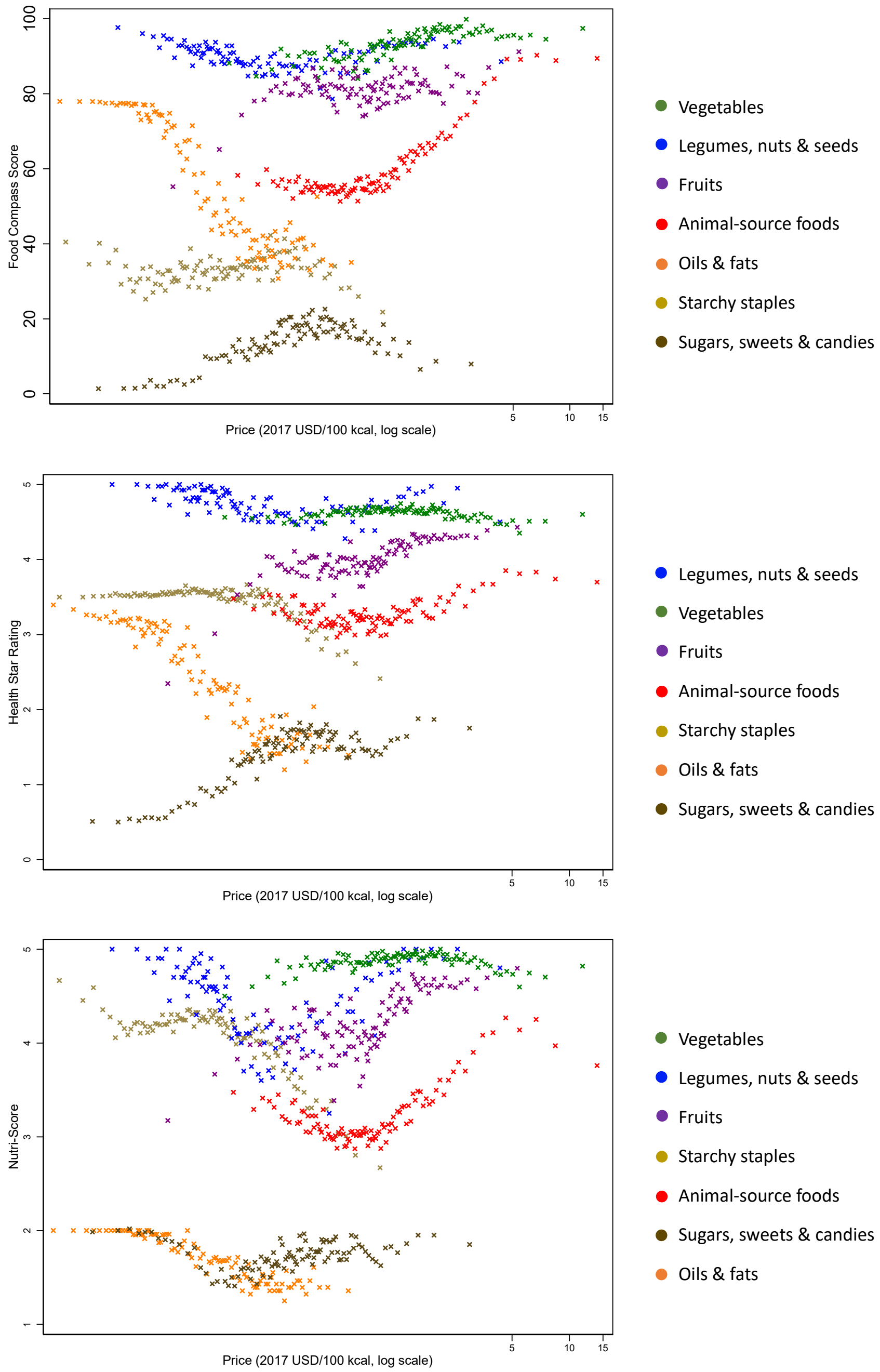

*Note: Food Compass Score, Health Star Rating, and Nutri-Score calculations from Barrett et al. (2024) matched to average retail food prices from the World Bank International Comparison Program in 2011 and 2017 for 811 food items in 181 countries. Price in 2017 USD per kilocalorie is shown in natural-log scale. Figures are binned scatter plots, where each food group is represented by 100 data points, each of which is the mean value of the y-axis variable at the mean level of price per 100kcal across 100 equal-sized bins of price per 100kcal.*

**Supplementary Figure 4b. Estimated mean Food Compass Score, Health Star Rating, and Nutri-Score conditional on price per kilogram, by food group**

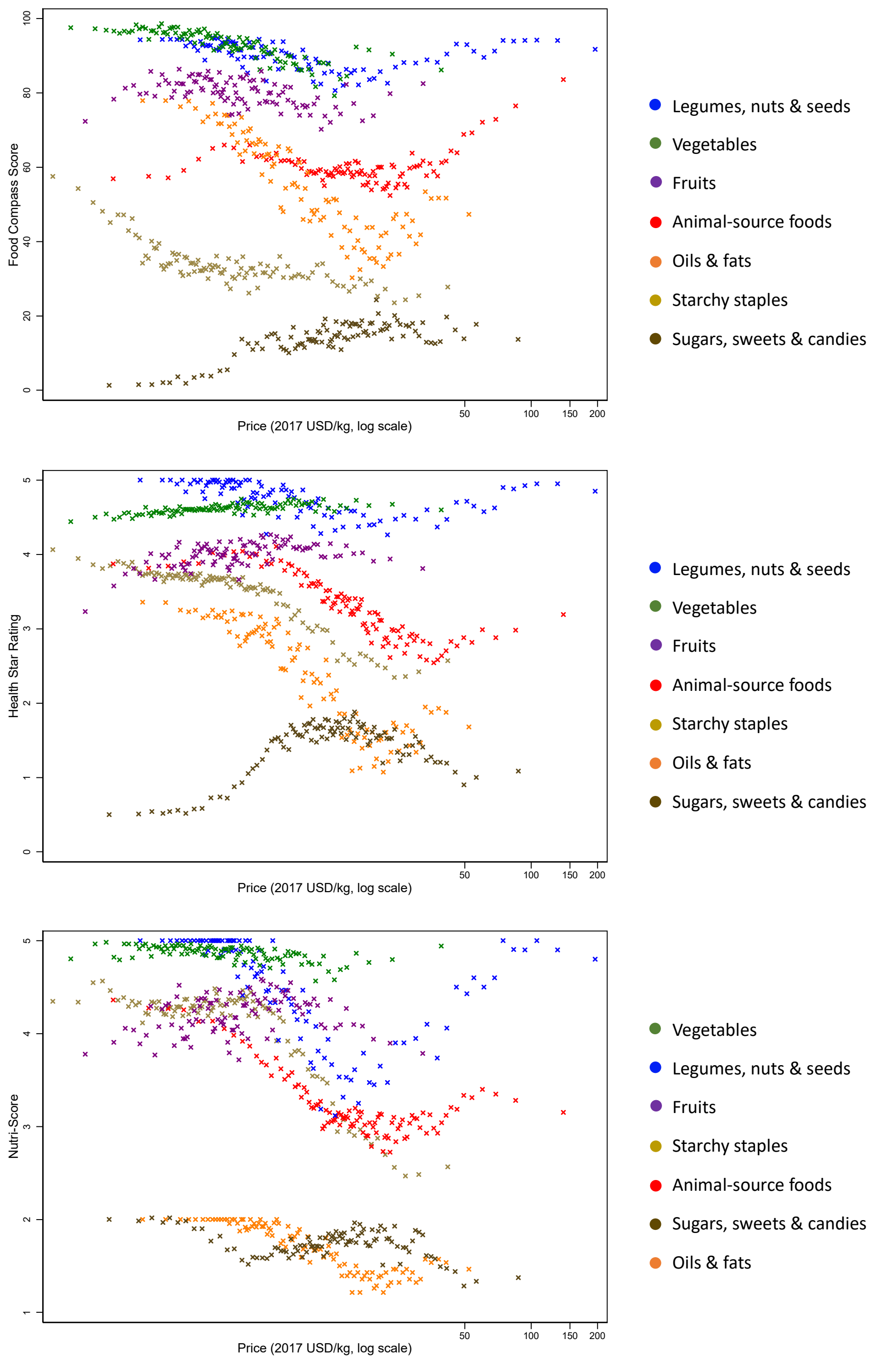

*Note: Food Compass Score, Health Star Rating, and Nutri-Score calculations from Barrett et al. (2024) matched to average retail food prices from the World Bank International Comparison Program in 2011 and 2017 for 818 food items in 181 countries. Price in 2017 USD per kilogram is shown in natural-log scale. Figures are binned scatter plots, where each food group is represented by 100 data points, each of which is the mean value of the y-axis variable at the mean level of price per kilogram across 100 equal-sized bins of price per kilogram.*

**Supplementary Figure 4c. Estimated mean Food Compass Score, Health Star Rating, and Nutri-Score conditional on price per recommended daily intake, by food group**

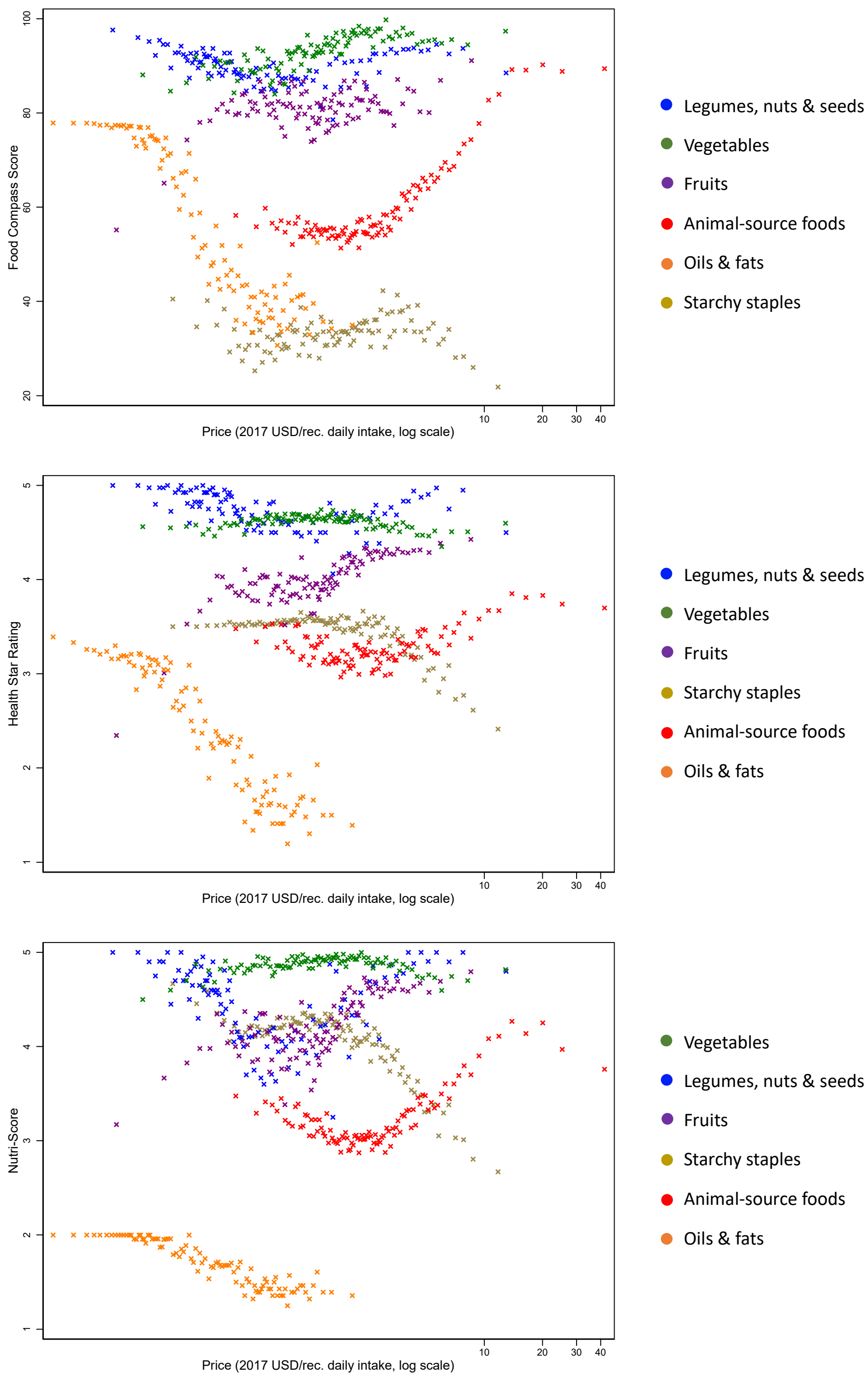

*Note: Food Compass Score, Health Star Rating, and Nutri-Score calculations from Barrett et al. (2024) matched to average retail food prices from the World Bank International Comparison Program in 2011 and 2017 for 707 food items in 181 countries. Price in 2017 USD per recommended daily intake is shown in natural-log scale. Estimates per recommended daily intake omit "sugars, sweets & candies" because there is no recommended intake of this food group. Figures are binned scatter plots, where each food group is represented by 100 data points, each of which is the mean value of the y-axis variable at the mean level of price per recommended daily intake across 100 equal-sized bins of price per recommended daily intake.*

**Supplementary Figure 4d. Associations between price per kilocalorie Food Compass Score, Health Star Rating, and Nutri-Score by food group**

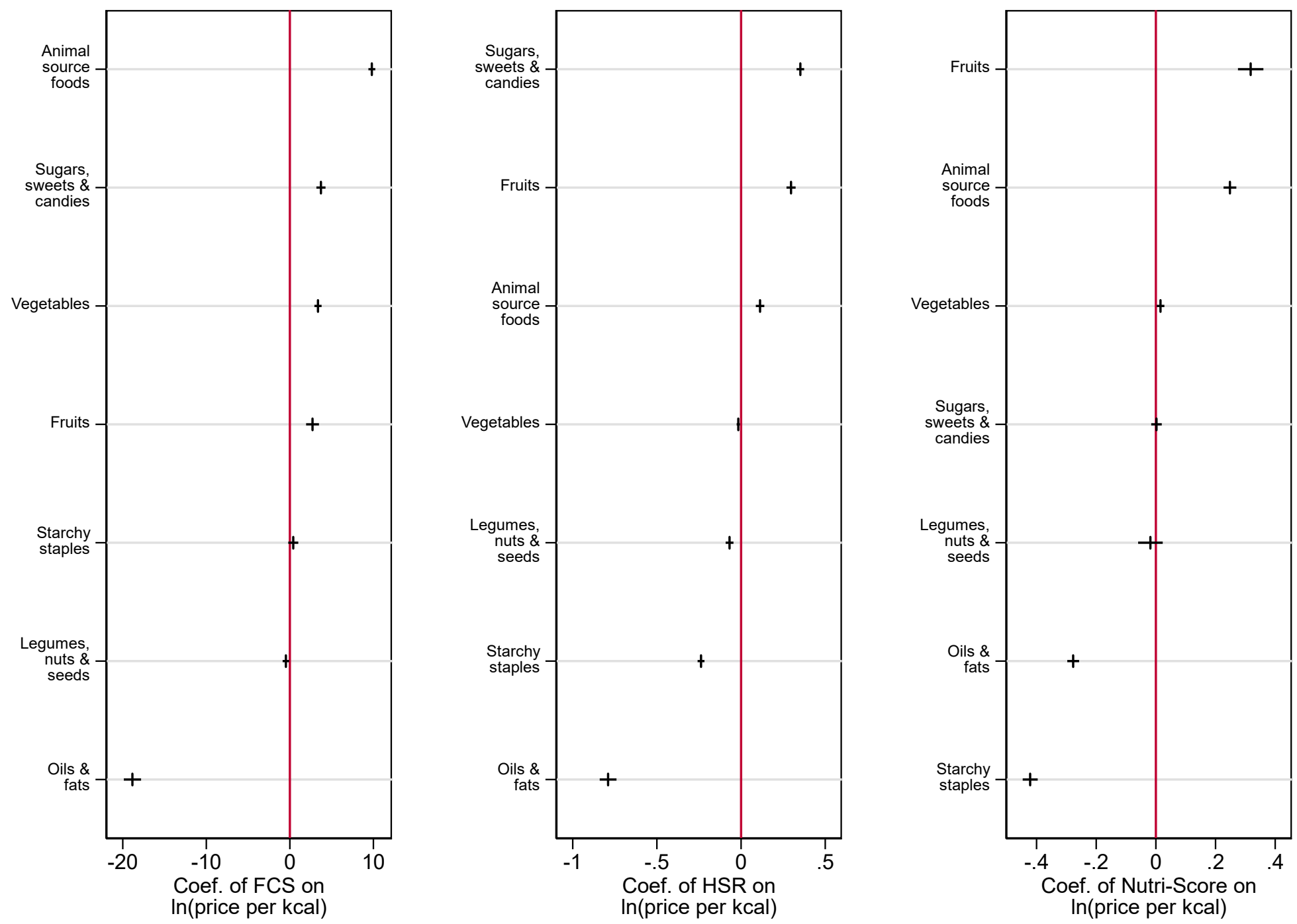

*Note: Tick marks represent coefficients and 95% confidence intervals of linear regressions of Food Compass Score (FCS), Health Star Rating (HSR), and Nutri-Score on log(price) with country fixed effects, stratified by food group.*

**Supplementary Figure 4e. Associations between price per kilogram Food Compass Score, Health Star Rating, and Nutri-Score by food group**

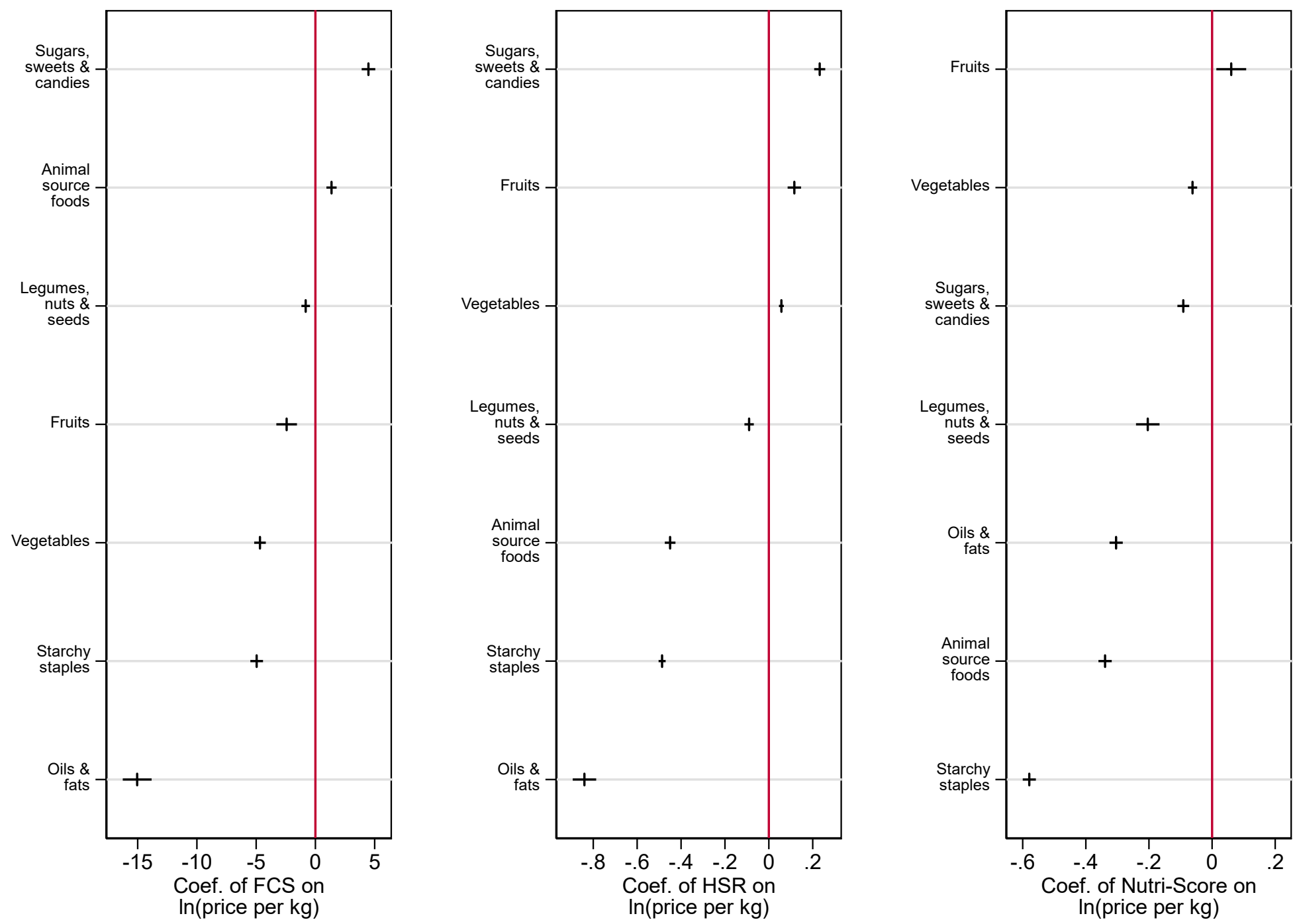

*Note: Tick marks represent coefficients and 95% confidence intervals of linear regressions of Food Compass Score (FCS), Health Star Rating (HSR), and Nutri-Score on log(price) with country fixed effects, stratified by food group.*

**Supplementary Figure 4f. Associations between price per kilocalorie Food Compass Score, Health Star Rating, and Nutri-Score by food group**

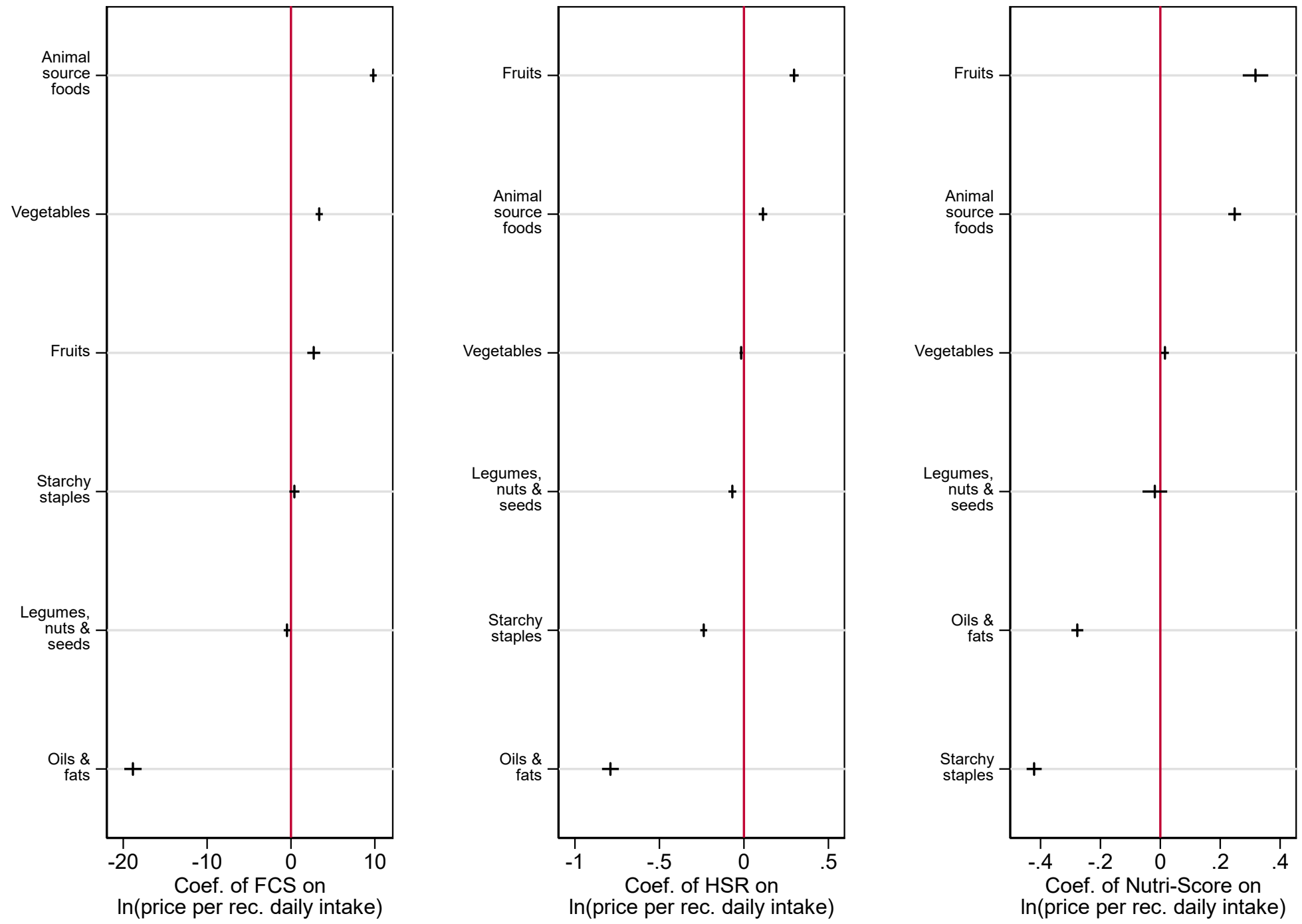

*Note: Tick marks represent coefficients and 95% confidence intervals of linear regressions of Food Compass Score (FCS), Health Star Rating (HSR), and Nutri-Score on log(price) with country fixed effects, stratified by food group. Estimates per recommended daily intake omit "sugars, sweets & candies" because there is no recommended intake of this food group.*

**Supplementary Figure 4g. Food Compass Score, Health Star Rating, and NutriScore by deciles of price per kilocalorie and food group**

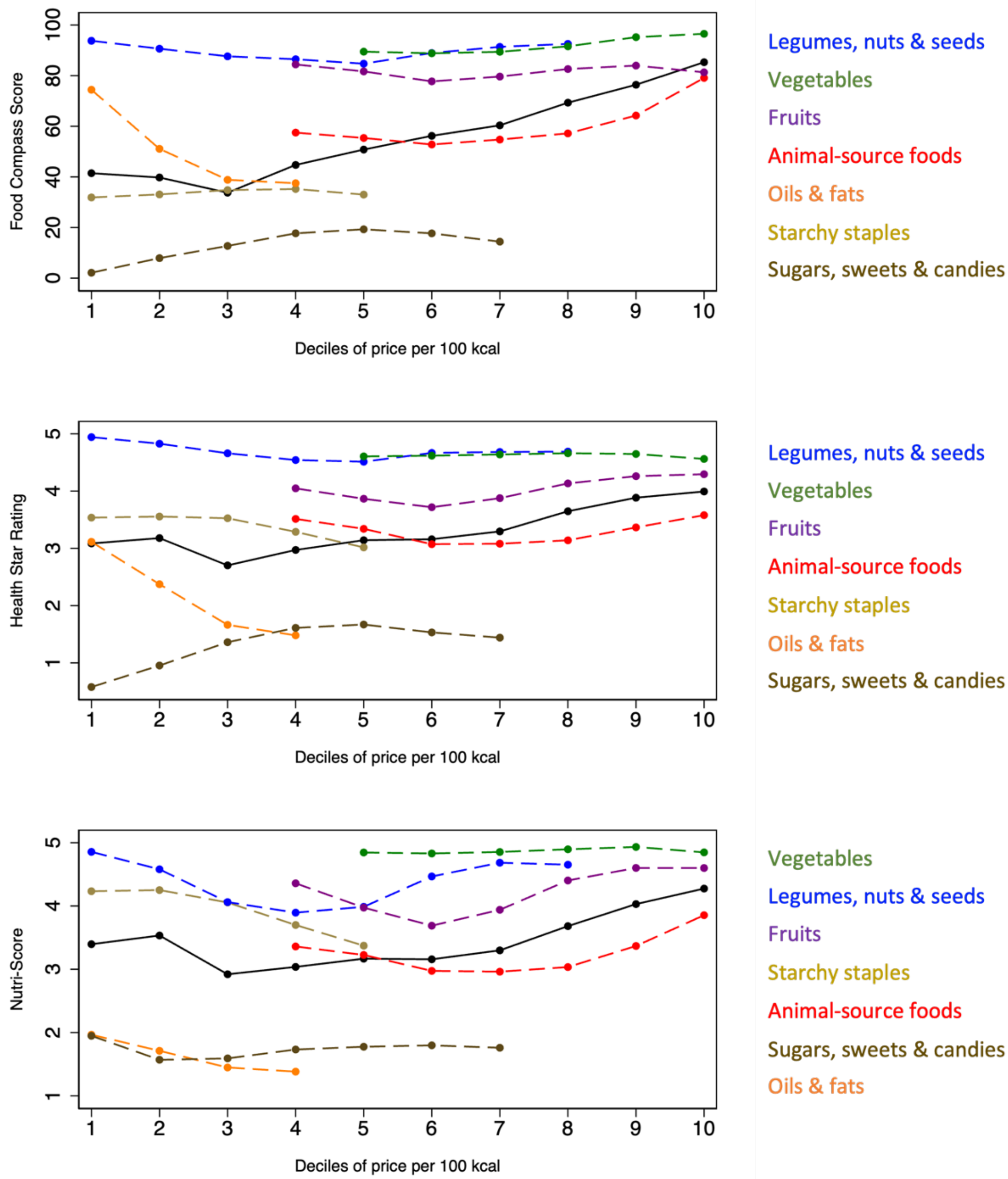

*Note: Food Compass Score, Health Star Rating, and Nutri-Score calculations from Barrett et al. (2024) matched to average retail food prices from the World Bank International Comparison Program in 2011 and 2017 for 799 food items in 181 countries. Deciles that contain less than 5 percent of the observations for a food group are omitted.*

**Supplementary Figure 4h. Food Compass Score, Health Star Rating, and NutriScore by deciles of price per kilogram and food group**

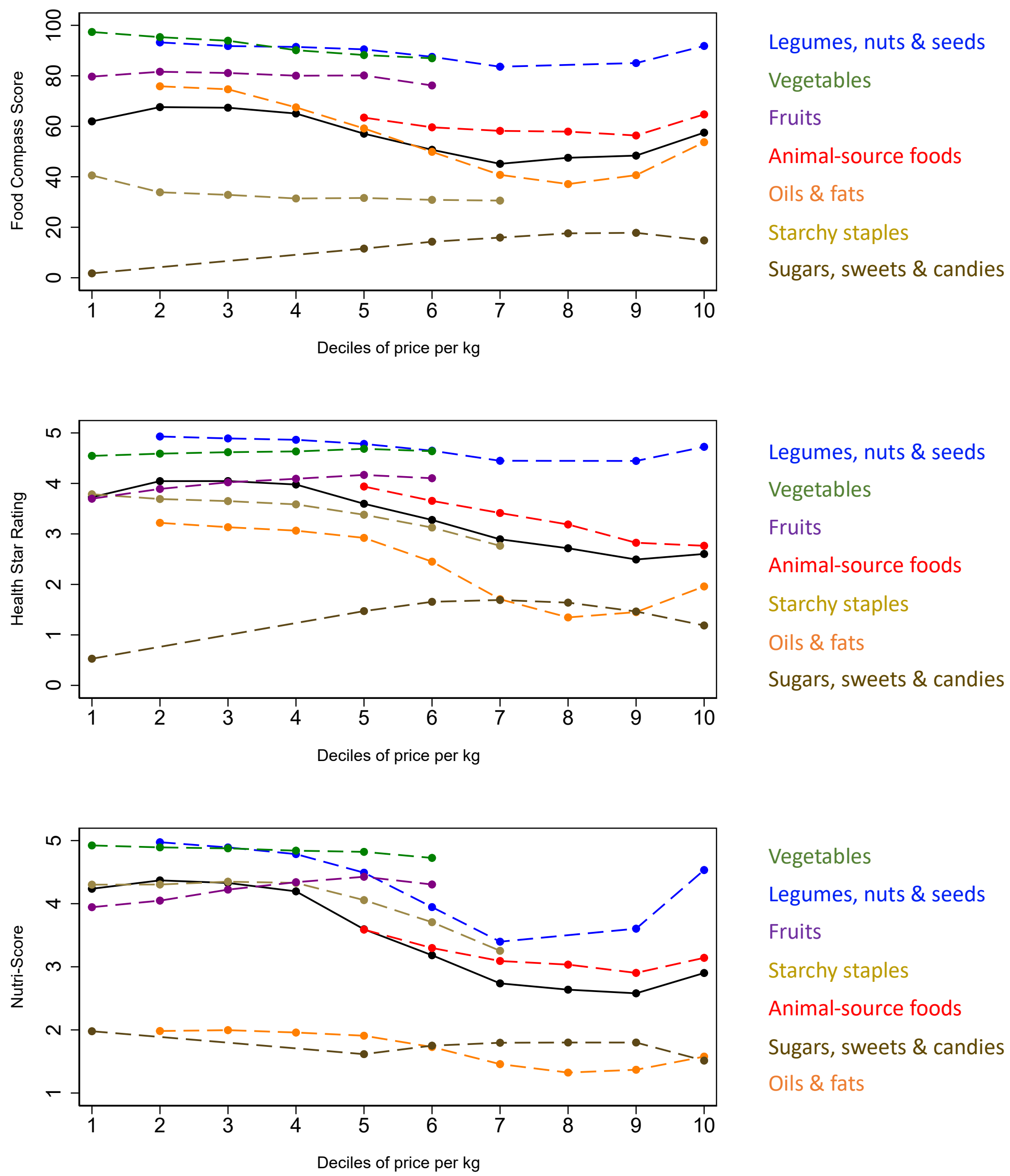

*Note: Food Compass Score, Health Star Rating, and Nutri-Score calculations from Barrett et al. (2024) matched to average retail food prices from the World Bank International Comparison Program in 2011 and 2017 for 793 food items in 181 countries. Deciles that contain less than 5 percent of the observations for a food group are omitted.*

**Supplementary Figure 4i. Food Compass Score, Health Star Rating, and NutriScore by deciles of price per recommended daily intake and food group**

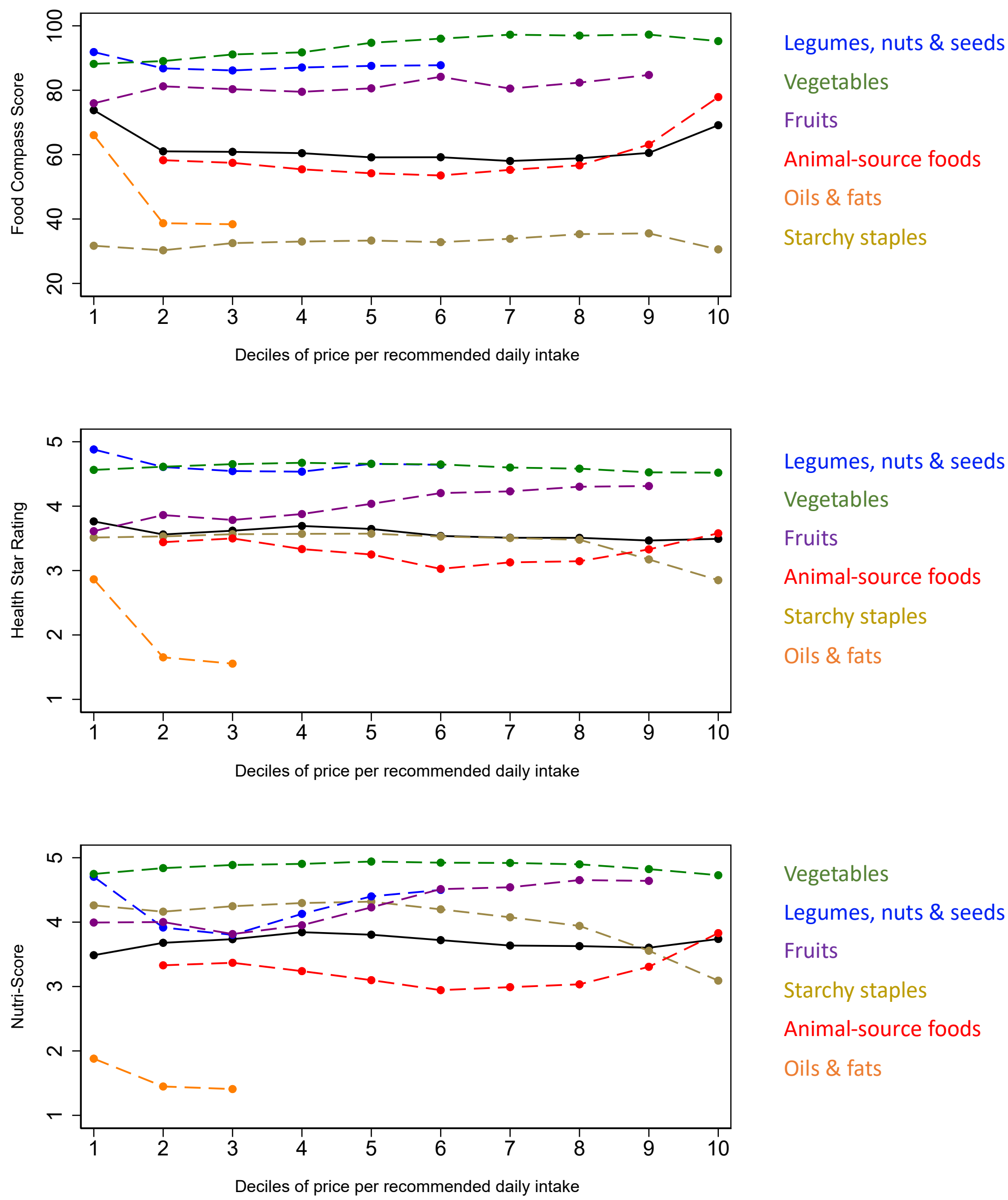

*Note: Food Compass Score, Health Star Rating, and Nutri-Score calculations from Barrett et al. (2024) matched to average retail food prices from the World Bank International Comparison Program in 2011 and 2017 for 698 food items in 181 countries. Estimates per recommended daily intake omit "sugars, sweets & candies" because there is no recommended intake of this food group. Deciles that contain less than 5 percent of the observations for a food group are omitted.*

## Appendix 5. Comparing retail food prices, carbon footprint, and water footprint, and nutrient profile per kilocalorie and per recommended daily intake in 2011 versus 2017

### Supplementary Figure 5a. Estimated mean carbon footprint, water footprint, and nutrient profile conditional on price per kilocalorie in 2011 and 2017, by food group

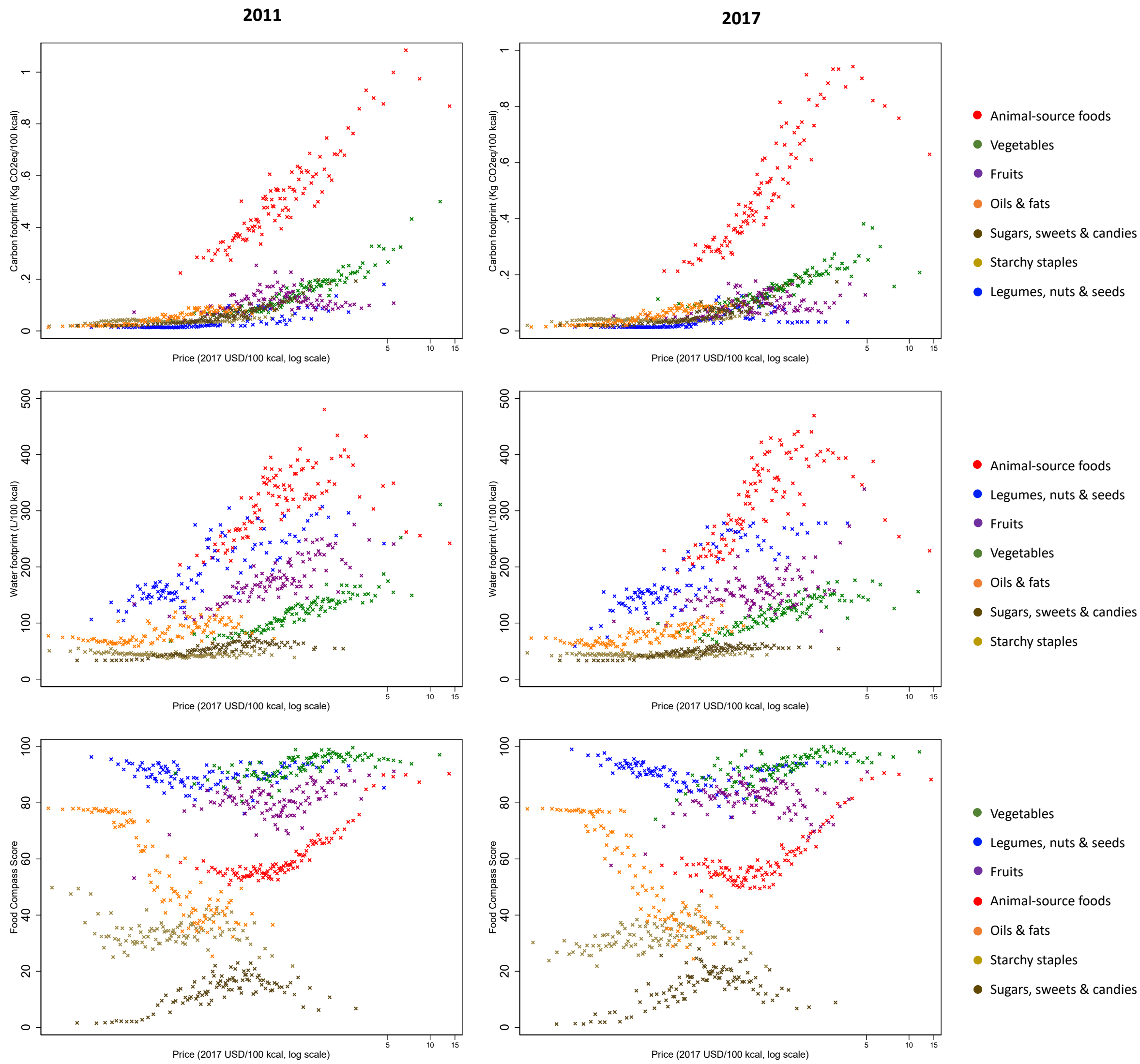

*Note: Carbon footprint and water footprint estimates from Petersson et al. (2021) and Food Compass Score calculations from Barrett et al. (2024) matched to average retail food prices from the World Bank International Comparison Program in 2011 and 2017 for 811 food items in 181 countries. Price in 2017 USD per kilocalorie is shown in natural-log scale. Figures are binned scatter plots, where each food group is represented by 100 data points, each of which is the mean value of the y-axis variable at the mean level of price per 100kcal across 100 equal-sized bins of price per 100kcal.*

**Supplementary Figure 5b. Estimated mean carbon footprint, water footprint, and nutrient profile conditional on price per recommended daily intake in 2011 and 2017, by food group**

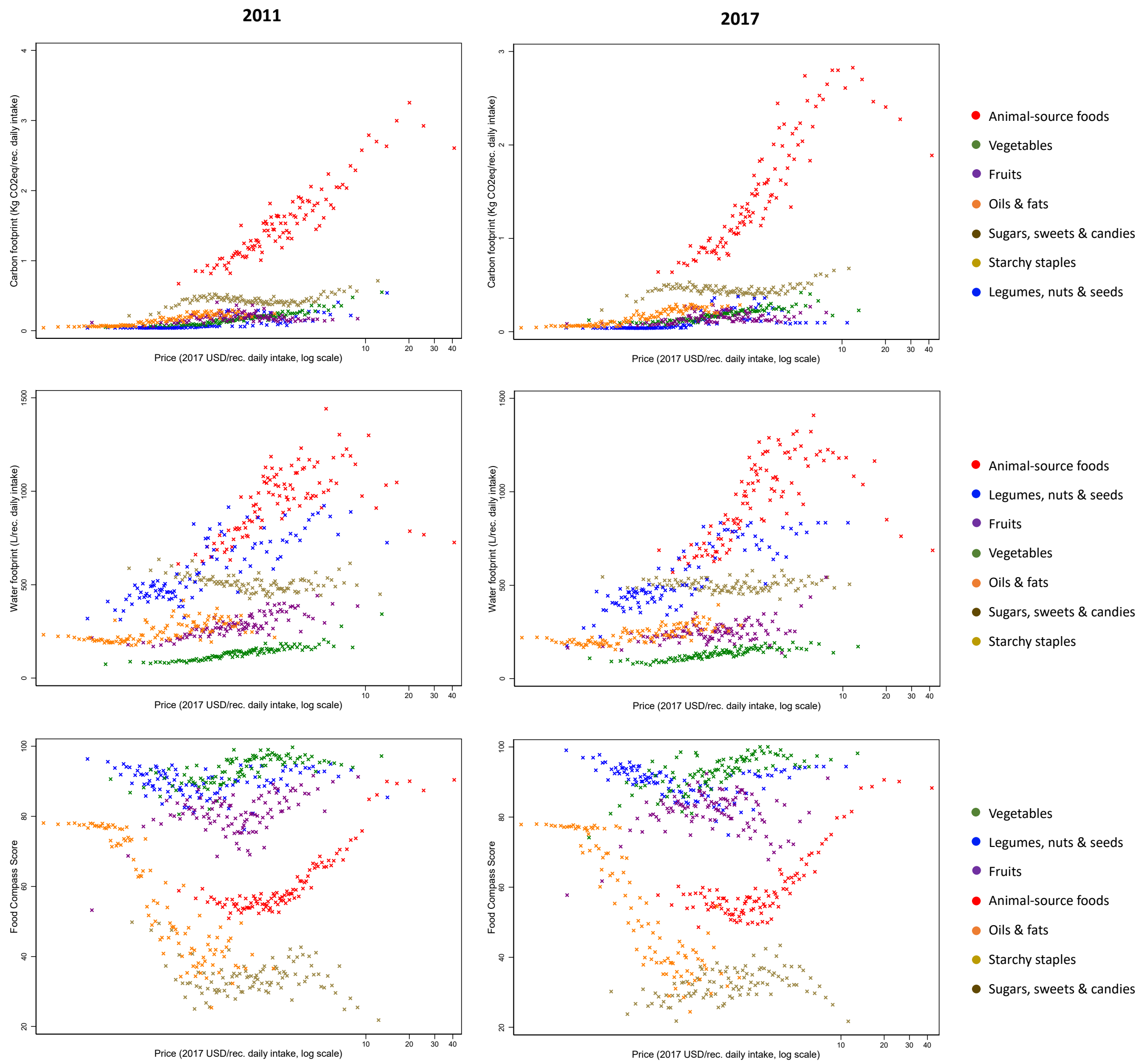

*Note: Carbon footprint and water footprint estimates from Petersson et al. (2021) and Food Compass Score calculations from Barrett et al. (2024) matched to average retail food prices from the World Bank International Comparison Program in 2011 and 2017 for 707 food items in 181 countries. Price in 2017 USD per recommended daily intake is shown in natural-log scale. Estimates per recommended daily intake omit "sugars, sweets & candies" because there is no recommended intake of this food group. Figures are binned scatter plots, where each food group is represented by 100 data points, each of which is the mean value of the y-axis variable at the mean level of price per recommended daily intake across 100 equal-sized bins of price per recommended daily intake.*